\documentclass[1p,authoryear, 12pt, review]{elsarticle}
\usepackage{lipsum}

\makeatletter
\def\ps@pprintTitle{%
 \let\@oddhead\@empty
 \let\@evenhead\@empty
 \def\@oddfoot{}%
 \let\@evenfoot\@oddfoot}

\usepackage{fancyhdr}

\usepackage{setspace}
\usepackage{graphicx}
\usepackage{bm}
\usepackage{amsmath}

\usepackage{natbib}
\usepackage[footnotesize]{caption}

\usepackage{subcaption}
\usepackage{float}
\usepackage{multirow}
\usepackage[flushleft]{threeparttable}
\usepackage{booktabs}
\usepackage[splitrule]{footmisc} 

\usepackage{lscape}
\usepackage{makecell}
\usepackage{longtable}

\usepackage{hyperref}
\usepackage{xcolor}
\hypersetup{
	colorlinks,
	linkcolor={blue!50!black},
	citecolor={blue!50!black},
	urlcolor={blue!80!black},
	breaklinks=true
} 

\usepackage[nameinlink,capitalise]{cleveref}
\Crefname{figure}{Fig.}{Figs.}
\Crefname{equation}{Eq.}{Eqs.}

\usepackage{amssymb}
\usepackage{amsmath}

\usepackage{tabulary}
\usepackage{tabularx}
\newcolumntype{Y}{>{\raggedleft\arraybackslash}X}
\usepackage{filecontents}
\usepackage{float}
\usepackage{longtable}
\usepackage{pdfpages}
\usepackage{booktabs}
\usepackage{fullpage}
\usepackage{verbatim}
\usepackage{lscape}
\usepackage{soul}
\numberwithin{equation}{section}
\usepackage[font=normalsize, justification=justified,singlelinecheck=false]{caption}
\usepackage{setspace}
\usepackage{lineno}
\usepackage{geometry}
\usepackage{color}
\usepackage{natbib}

\begin{document}

\begin{frontmatter}
\title{\bf{Extremal Dependence in Australian Electricity Markets}}

\author{Lin Han}
\ead{lin.han@mq.edu.au}
\address{Department of Actuarial Studies and Business Analytics, Macquarie University}

\author{Ivor Cribben}
\ead{cribben@ualberta.ca}
\address{Department of Accounting and Business Analytics, University of Alberta}

\author{Stefan Tr\"uck}
\ead{stefan.trueck@mq.edu.au}
\address{Department of Actuarial Studies and Business Analytics, Macquarie University}

\setstretch{1}

\begin{abstract}

Electricity markets are significantly more volatile than other comparable financial or commodity markets. Extreme price outcomes and their transmission between regions pose significant risks for market participants. We examine the dependence between extreme spot price outcomes in the Australian National Electricity Market (NEM). We investigate extremal dependence both in a univariate and multivariate setting, applying the extremogram developed by \cite{davis2009} and \cite{davis2011,davis2012}. We measure the persistence of extreme prices within individual regional markets and the transmission of extreme prices across different regions. With both 5-minute and 30-minute price data, we find that extreme prices are more persistent in the market with a higher share of intermittent renewable energy. We also find that persistence of extreme prices is more prevalent in more concentrated markets. We also show significant extremal price dependence between different regions, which is typically stronger between physically interconnected markets. The dependence structure of extreme prices shows asymmetric and time-dependent patterns. Applying the extremograms, we further show the effectiveness of the Australian Energy Market Commission's 2016 rebidding rule with respect to reducing the share of isolated price spikes that are often considered as an indication of strategic bidding. Our results provide important information for hedging decisions of market participants and for policy makers who aim to reduce market volatility and extreme price outcomes through effective regulations which guide the trading behaviour of market participants as well as improved network interconnections.

\end{abstract}

\begin{keyword}
 Electricity Prices \sep Extreme Prices \sep Dependence \sep  Extremogram  \sep  Interconnectors  \sep  Strategic Behaviours

\JEL{G11,  G12,  G13}
\end{keyword}

\end{frontmatter}

\newpage


\section{Introduction}\label{intro}

Since the 1990s, the global power sector has been going through a significant period of restructuring and deregulation. Similar to other commodities, electricity in many countries is now traded in competitive markets with both spot and derivative contracts \citep{harris2011electricity}. Despite recent advances in energy storage such as large scale batteries, electricity is still widely considered as a non-storable good which requires immediate delivery after generation. The non-storable nature of electricity under competitive rules of the market causes an extremely tight relationship between demand and supply. As a result, relatively small fluctuations in generation or transmission can cause significant changes in electricity prices within a short time interval, which appear as short-lived but extreme spikes \citep{Weron2006,clements2015modelling,manner2016modeling}. This leads electricity spot prices to be substantially more volatile than those of comparable commodities or financial instruments.
The extreme outcomes in electricity markets are of great concern to market participants and can be detrimental for businesses (e.g. large retailers who purchase electricity with spot prices but sell electricity to customers with a fixed price) without an appropriate hedging strategy.
In addition, in a multi-regional context where interconnection is in place, the extreme outcomes often transmit to connected markets through interregional trade of electricity.
The possibility of joint price spikes in different regions is of particular concern to participants that operate businesses in several regional markets.

This paper examines the dependence between extreme price events in the spot electricity market in the Australian National Electricity Market (NEM).
The NEM operates as a wholesale spot electricity market where supply and demand are matched in real time. It is an interconnected grid across five regional market jurisdictions: New South Wales (NSW), Queensland (QLD),  South Australia (SA), Tasmania (TAS) and Victoria (VIC).
Within the NEM, electricity can be transmitted between adjoining regions via high voltage transmission lines, so-called interconnectors.
This typically happens when the electricity price in another region is sufficiently low to displace the local supply, or when the local power demand exceeds the available local generation capacity. The market interconnection allows electricity to be traded from a low price area to a high price area to mitigate the price difference, limit the market power of local generators in separate regions, improve the reliability of power supply, and reduce overall electricity costs to customers \citep{agpc13}. However, the extent of these benefits is limited by the capacity of interconnectors.
As raised by \cite{garnaut11} and the \cite{agpc13}, a potential underinvestment in interconnectors may have led the NEM to remain relatively isolated \citep{HiggsH2009,apergis2016integration,nepal2013testing} with substantial price difference between regions and the market power retained within a small number of generators in each state.
Interestingly, despite the occasions when substantial price differences are observed, there have been other periods when extreme prices occurred simultaneously or with a certain time lag in several regional markets \citep{IgnTr:2016}.
Therefore, there is a strong interest for market participants in the NEM to analyse the complex dependence structure between extreme events of electricity prices.

This study aims to provide a better understanding of extreme price outcomes in electricity markets as well as the relationship between extreme events both within individual regional electricity spot markets and across different regions.
We present one of the first applications of the extremogram developed by \cite{davis2009} and \cite{davis2011,davis2012} to spot electricity prices to thoroughly address this complex relationship.

In addition, as discussed by \cite{clements2016strategic}, a potentially important factor that contributes to large price spikes and raises increasing concerns from both the academic literature and market regulators is the strategic behaviour of generators.
In the Australian NEM where half-hourly settlement prices are determined as the average of six 5-minute dispatch intervals, it is possible for generators to implement profit-maximising strategies: for example, generators can create a price spike by withholding capacity in one dispatch interval and adding capacity with low prices afterwards to ensure dispatch, and receive a high average price overall \citep{dungey2018strategic}. The \cite{australian2015final} reported a positive relationship between extreme prices and the number of late rebids shifting the offered capacity to higher prices. \cite{hurn2016smooth} in particular attribute price spikes in QLD to the strategic behaviour of generators. Due to concerns about the negative impacts of this strategic bidding behaviour, new market rules have been implemented to ensure generators bid in good faith and are subject to reviews and amendments, see, for example, \cite{australian2015final}.
Since strategic bidding of market participants is highly relevant to the occurrence of extreme prices and the interregional price difference \citep{clements2016strategic}, our study, by examining empirical extremograms before and after a particular market rule change, also aims to shed light on the impacts of market regulation on the behaviour of market participants and efficient market pricing.

The extremogram is a flexible measure which quantifies the dependence of extreme events. In a univariate setting, the extremogram can be considered as an extreme-value analog of the widely used autocorrelation function (ACF) for a stationary time series. While the ACF typically has limited value when heavy-tailed nonlinear processes are considered, the extremogram is particularly designed to measure the dependence between tail events \citep{davis2009}. In particular, the extremogram computes the extent to which an extreme value of a time series has an impact on a future value of the same time series (i.e. univariate extremogram) or another time series (i.e. cross-extremogram).
The estimation of the extremogram fits the time series data of electricity spot prices since these processes exhibit both heavy tails (i.e. extreme values) and volatility clustering (i.e. persistent spikes) over time.
Meanwhile, as the extremogram only considers observations at tails which are extreme and rare by nature, a large sample size is required for accurate estimation of the extremograms. Thus, the long time series data available and the high frequency of electricity spot prices are advantageous for the application of extremograms.
With the application of extremograms, the persistence and transmission of extreme events can be described both quantitatively and graphically.

We apply the extremograms to all five state-based regional markets in the Australian NEM. Using high frequency spot electricity price data with both half-hourly and 5-minute frequencies, we examine the persistence and transmission of extreme electricity prices in the NEM for a sample period of nine years from July 1, 2009 to June 30, 2018.

Estimating a univariate extremogram to each regional market, we find highest persistence of extreme prices in SA, the market with a higher share of intermittent renewable generation.
We also find high persistence of extreme prices in QLD, which is a more concentrated market, and in TAS, which is the smallest and most distinct market from the mainland in the NEM. On the other hand, NSW and VIC exhibit substantially lower persistence of extreme prices.
In each region, significant periodicity is observed in the estimated extremograms, i.e. given an existing price spike, the highest likelihood for another spike is found around the same time on the next day.
Concerning the estimated cross-extremograms, which measure the extremal dependence between different markets, we find the strongest extremal dependence between adjacent markets that are physically interconnected, for example, NSW and QLD, NSW and VIC, and SA and VIC. This indicates a relatively high likelihood of joint price spikes or contagion effects for these pairs of markets.
On the other hand, significantly lower extremal dependence is found between markets that are geographically distant and not directly interconnected, such as NSW and TAS, and SA and TAS. Interestingly, despite the existing \emph{Basslink} interconnector which is a submarine cable between TAS and VIC, the estimated extremal dependence between the regional prices in these two markets is relatively low. These transmission effects vary with different forward-looking intervals (i.e. the lags used in the extremogram calculation) and are asymmetric between regions.

Furthermore, we conduct a case study by estimating the extremogram in order to investigate the impact of a particular regulatory market event -- the Australian Energy Market Commission (AEMC)'s rebidding rule change \citep{australian2015final} around `last minute rebids' which became effective from 1 July 2016. This change in regulation is aimed to prevent false and misleading bidding behaviour. Overall, our results suggest that the rebidding rule change was effective in reducing strategic bidding behaviour among market participants.

Our study contributes to the literature in three dimensions. First, we apply the extremogram method and confirm its effectiveness in capturing the transmission of extreme electricity prices. Second, compared with the existing literature, we provide a more in-depth analysis of extremal dependence in the Australian NEM by estimating more detailed patterns of these effects. Evidence such as the magnitudes, directions and persistence of extremal dependence is documented. Third, our results provide empirical evidence on the effectiveness of the AEMC's 2016 rebidding rule change, which has not been studied yet in the existing literature.

We provide important insights for both market participants and regulators in electricity markets. 
Our results on extremal dependence across regions provide important information on interregional risk management and hedging strategies. This is of particular interest to traders and merchant interconnectors who make a profit by purchasing electricity from a region where the electricity price is relatively cheap and selling it to a region where the local price is high.
Our findings regarding the persistence of extreme prices also have important implications for the operation of large industrial customers with load curtailment arrangements or battery storage, because the timing of buying and selling power (i.e. the ability to buy/use electricity during low price periods and sell/stop using it during high price periods) is highly relevant to their revenues.

 The analysis of extreme price persistence is also important for peak load generators such as gas fired plants which require information on future spikes, since peak load generators typically start generating only when there is a high probability of more price spikes in upcoming half-hourly trading intervals.
Finally, from a regulatory perspective, our findings on the extremal dependence structure for different regional markets may be helpful for decisions on constructing new interconnectors and developing policies to guide trading strategies of market participants. Such decisions are important in order to achieve a more integrated market, limit the market power of local generators and reduce the overall price volatility in the NEM.

The remainder of this study is organised as follows. \Cref{literature} provides an overview of the literature related to our study. \Cref{facts} briefly reviews the institutional structure of the Australian NEM. \Cref{extremo} introduces the applied extremogram method, while \Cref{empirical} presents our empirical results. \Cref{section: rebidding casestudy} examines the impact of the rebidding rule change by the AEMC on the persistence of price spikes, using the extremogram. Finally, \Cref{conclusions} discusses the main findings of this study and concludes.

\section{Related Literature}\label{literature}

The worldwide deregulation of electricity markets since the 1990s has fostered a rapidly growing literature aiming at modelling and forecasting electricity prices. However, only recently a number of studies focus particularly on extreme events or spikes in electricity prices, most of which are conducted in a univariate context. Overall, clustering and persistence are important features of extreme events in electricity markets that are documented in the previous literature. Various models are applied to capture these features.
In particular, \cite{christensen2009never} propose a poisson autoregressive framework to model electricity price spikes. They argue that persistence is an important feature of spikes to be accounted for.
\cite{christensen2012forecasting} use an autoregressive conditional hazard model to forecast both the occurrence and size of spikes.
\cite{korniichuk2012forecasting} employs a copula model to capture the dependence in magnitudes of extreme electricity prices and a negative binomial duration model to forecast the occurrence of extreme prices.
\cite{clements2013semi} propose a semi-parametric approach which is superior in capturing the nonlinearity of the spiking process, while
\cite{eichler2014models} use dynamic logit models to forecast the occurrence of electricity price spikes. They show that  spikes often occur in clusters and tend to reappear at the same time on the following day.
Furthermore, \cite{herrera2014modeling} propose a self-exciting marked point process approach based on either durations or intensities to model spikes.

So far the analysis of electricity markets in a multivariate context is still considered as limited despite some recent growth. The focus is typically on measuring the overall interdependence, co-movement and integration level between interconnected markets.
\cite{devany99}, \cite{de1999price}, \cite{park06} and \cite{Dempster08} examine the market integration level in US spot electricity markets, using vector autoregressive (VAR) based models and cointegration analysis. \cite{devany99,de1999price} find evidence of a highly integrated and efficient power market in the western US, while \cite{Dempster08} suggest only a moderate integration level in the same market. \cite{park06} find a interrelationship structure between eastern and western US electricity markets that is dependent on the considered length of time horizon.
\cite{haldrup06} develop a Markov Regime Switching fractional integration model and find time-varying patterns in the price dependence between regional markets in the Nordpool market in Europe.
\cite{zachmann08} investigates the integration level of European electricity markets. Applying principal component analysis, the findings of this study reject the hypothesis of an integrated market. However, pairwise price convergence is observed between several countries when congestion costs are considered.
\cite{lepen08} estimate the return and volatility spillover effects in three major electricity forward markets in Europe through a VAR-BEKK model.
More recently, \cite{ciarreta2015analysis} use multivariate GARCH models to estimate mean and volatility spillovers of prices among European electricity markets. They find evidence that increasing interconnections and homogeneous rules of market operations contribute to higher levels of price convergence and market integration.
\cite{de2015germany,de2014reassessing} employ fractional cointegration methods to assess the price convergence between different European electricity markets, because electricity spot prices in the considered markets are found to be intermediate between a non-stationary and a stationary process.
\cite{muniain2020probabilistic} use bivariate data to forecast electricity prices. A bivariate jump component is included in the model which improves the performance in forecasting correlated time series.

In the Australian context, \cite{WKH2005} and \cite{HiggsH2009} employ a family of multivariate GARCH and constant and dynamic conditional correlation (i.e. CCC and DCC) type models to examine volatility spillovers across regional electricity markets. They find significantly higher conditional correlations between well-connected markets and weaker interdependence between the least interconnected markets.
\cite{smith2010}, \cite{aderounmu2014modeling}, \cite{smith2015copula}, \cite{IgnTr:2016} and \cite{smith2018econometric} apply various copula models to examine the nonlinear dependence structure and tail dependence between regional electricity spot prices in Australia.
Furthermore, two studies investigate the degree of market integration in Australian electricity markets. In particular, \cite{nepal2013testing} conduct pairwise unit root tests, a cointegration analysis, and a time-varying coefficient model based on the Kalman filter to examine the integration level of interconnected regional markets in the Australian NEM. Their results suggest that full integration has not been achieved due to the limitation of interconnector capacities.
\cite{apergis2016integration} test for price convergence across different electricity markets in Australia by employing a clustering groups approach. Considering the five regional markets in the NEM as well as the Western Australia (WA) market, they find long-term price convergence in three groups: NSW, QLD and VIC; SA; and TAS and WA. The separation of these groups is influenced by factors such as physical interconnections, market structures and competition levels. \cite{han2020volatility} examine volatility spillover effects across regional Australian spot electricity, suggesting that spillovers are significantly influenced by regional proximity and interconnectors. \cite{yan2020} apply dynamic network analysis to a system of regional electricity markets. Their results suggest that South Australia is the most vulnerable market, while Victoria is the most influential one. Interestingly, the developed network measures also exhibit some predictive power for spot price behaviour.

Among studies focusing on the multivariate analysis of spot electricity markets, only a few recent studies focus on extreme price events. In particular, \cite{clements2015modelling} assess the transmission of price spikes across interconnected regional markets in the Australian NEM by developing a multivariate self-exciting point process model. They find that the size of price spikes is impacted by the available interconnector capacity: the size of spikes tends to be smaller where there is greater excess capacity into a region. \cite{frolova} applies various statistical methods, including quantile regression, the cross-quantilogram, the extremogram and cross-extremogram, as well as Gumbel copulas to examine extreme value dependence between different electricity markets. \cite{bigerna2017renewables} apply a contagion model on different electricity markets in Italy. They separate between contagion arising only in exceptional market conditions and the interdependence between markets.
Dynamic copula models are developed in \cite{manner2016modeling,manner2019forecasting} to forecast the occurrence of spikes in multiple regional markets.
\cite{do2020interconnectedness} conduct a higher moment interconnectedness analysis in the Australian NEM and suggest that the transmission of extreme price risk can be reduced by increasing transmission and generation capacity in the NEM.

Despite the limited focus on electricity price spikes in the multivariate context, existing studies typically put more effort on modelling or forecasting the occurrence and size of spikes. There is a lack of documented evidence on describing the characteristics and dynamics of these extreme price outcomes, for example, how persistent they are and the directions and timing of their transmissions, both in individual and interconnected markets. In addition, none of these studies with the exception of \cite{frolova} has applied the extremogram \citep{davis2009,davis2011,davis2012} to examine extremal dependence or use this technique to examine the impacts of market events.

A number of studies have recently developed methods to analyse dependence in quantiles of distributions of time series data, including, for example, quantilograms \citep{linton2007quantilogram} and cross-quantilograms \citep{han2016cross,todorova2017intraday}, quantile correlations \citep{schmitt2015quantile}, quantile autocorrelation functions \citep{li2015quantile}, and multivariate multi-quantile models \citep{white2015var}.
In comparison to these methods, the extremogram focuses on extreme quantiles. The extremogram method was first introduced by \cite{davis2009} and \cite{davis2011,davis2012} to measure serial dependence of extreme values in strictly stationary time series. Various aspects of extremograms were later discussed, including the asymptotic theory \citep{davis2013measures,cho2016asymptotic,matsui2016extremogram} and permutation methods to construct confidence bands for extremograms \citep{davis2012,drees2015bootstrapping}.
It has been recently applied to measure serial dependence of extreme observations and co-movements in stock markets \citep{bollerslev2013jump,herrera2018point,hautsch2020multivariate}, credit default swap (CDS) markets \citep{cont2011statistical} and extreme rainfall pattern identification \citep{rinaldi2018identification}. Furthermore, apart from using extremograms to measure the extremal dependence, \cite{mikosch2015integrated} implement a goodness of fit test in the frequency domain to conduct model selection based on the ability of extremograms to explain the dependence in tails. In our context, the application of the extremogram allows us to complement and extend empirical work focusing on the extremal dependence between spot electricity prices.

\section{The Australian National Electricity Market}\label{facts}

Electricity markets in Australia have undergone substantial changes since the 1990s \citep{AEMO2014}. Prior to 1997, the power system in Australia consisted of several vertically integrated and government owned businesses in each state. There was no interconnection between the individual states.
Electricity prices in the market were determined by the regulations of each state government to cover costs with potential required returns determined by the government.
With the objective of promoting power market efficiency, the Australian government commenced reform of the electricity markets in the late 1990s. In particular, the supply industry was separated into different segments, including transmission and distribution networks which remained in a regulated monopoly, and generation and retail markets which are now operated as competitive markets.
In addition, interconnection between adjoining states was introduced to form a national wholesale market for electricity trading.

The NEM started as a wholesale electricity market in Australia in December 1998.
It is now an interconnected grid including five state-based regional markets in NSW, QLD, SA, TAS and VIC.
Electricity trade in the NEM is established through a central pool where the outputs from all generators are aggregated and matched with the forecast demand.
The Australian Energy Market Operator (AEMO) manages this pool, following the National Electricity Law and relevant rules made by the Australian Energy Market Commission (AEMC). Unlike many electricity markets in the US and Europe, the NEM is not operated as a day-ahead market but is a spot market where electricity is traded in real time.
Note that two key terms regarding time intervals are important in the NEM context: the dispatch interval (5 minutes) and trading interval (30 minutes).
Generators in the NEM are required to submit offers and specify their available capacity and acceptable prices to cover their costs for the next half-hourly trading interval.
Every five minutes, AEMO operates a central dispatch process. It determines a price (i.e. dispatch price) for each region and the generators to be dispatched for the next five minutes based on the currently available offers and forecast demand through a least cost optimisation algorithm.
In particular, generators with lower offered prices get higher priority to be dispatched. They are dispatched in ascending order of offers until the market demand is met, and the dispatch price is determined by the highest offered price of all generators that get dispatched.
The final price (i.e. trading price) is set every half-hour for each region by averaging six 5-minute dispatch prices in a half-hourly trading interval \citep{AEMO2014}.
Generated electricity is settled each half-hour and generators in the same region are paid uniformly for their dispatched output at the regional spot price, which equals the half-hourly trading price in a region.
One critical feature of the NEM bidding mechanism is that generators are allowed to make variations to their existing bids at any time until five minutes before the dispatch interval. This feature aims to help achieve a high adaptability on the supply side to effectively respond to real-time grid or market changes. However, as argued by \cite{clements2016strategic} and \cite{dungey2018strategic}, this flexible rebidding mechanism also provides generators with the opportunity to submit strategic bids and rebids which can lead to undesirable consequences and detach the spot price from the fundamental costs of generation.

As pointed out by \cite{mayertrueck}, the Australian NEM is significantly more volatile and spike prone compared to other electricity markets in the world. The NEM also has an extremely high price cap (the maximum possible spot price) which is higher than price caps in other electricity markets around the world.
While the NEM spot price under normal conditions is typically below A\$100/MWh, the NEM price cap at the beginning of the sample period was A\$10,000/MWh on July 1, 2009, and increased to A\$14,200/MWh at the end of the sample period on June 30, 2018. During our sample period, there have been several occasions in each region when the spot price was close to or hit the specified market cap.
Therefore, hedging the risk induced by extreme price events is of particular importance for market participants in the NEM, since the surge of costs can be detrimental for businesses such as retailers who purchase electricity at the spot price and sell it at a fixed price to their customers.

Recent years have seen a rapid development of the market for electricity derivatives, including forward, futures and option contracts.
For example, the Australian Securities Exchange (ASX) provides various futures and options products for traders and market participants in the NEM to earn profits through trading or to hedge their risk according to particular needs\footnote{For available energy derivative contracts traded in the ASX, see the ASX website: https://www.asxenergy.com.au/.}.
Note that derivative contracts for electricity markets typically do not require physical delivery but are subject to financial settlement. As a result, participation in electricity derivatives markets does not require owning physical generation assets.
As well as monthly, quarterly and annual futures contracts that are priced based on the average spot prices of electricity during a certain delivery period, the ASX offers a number of alternative derivative contracts tailored towards extreme price outcomes. The most widely used product for extreme price hedging is the `Base Load Calendar Quarter \$300 Cap Futures' for each region \citep {ASX2019contract}. Note that for this product, `Base Load' hours are defined as the period from 00:00am to 24:00am on each calendar day. The payoff of these contracts depends on both the magnitude and the frequency of the extreme prices for a particular region in the calendar quarter that exceed A\$300/MWh. More specifically, the cash settlement price is calculated as $\frac{C-300\times D}{E}$, where $C$ is the sum of all spot prices for a certain region in the calendar quarter that exceed A\$300, $D$ is the total count of the occasions when the spot price is greater than A\$300/MWh, and $E$ is the total count of half-hourly intervals.
Estimating the payoff of such a contract is important for relevant market participants in the NEM to evaluate their hedging strategies. Since the calculation of payoffs requires the forecast of future spike count in a certain period, the persistence or potential recurrence possibility of a price spike would be highly relevant. Therefore, our extremogram analysis is of particular interest in terms of modelling the payoff distribution of these types of contracts in future work. 

The five regional markets in the NEM are currently linked through six interconnectors. In particular, two interconnectors (\textit{QNI} and \textit{Terranora}) are between NSW and QLD; two interconnectors (\textit{Heywood} and \textit{Murraylink}) are between VIC and SA; one \textit{Victoria-NSW Interconnector} is between NSW and VIC; and \textit{Basslink} is an undersea power cable between TAS and VIC \citep{AEMO2014}.
\textit{Basslink} is a merchant interconnector that derives profit by actively participating in the interregional trade of electricity, while the other five interconnectors are regulated interconnectors and receive fixed revenue based on the asset value regardless of their actual usage.
These interconnectors play a significant role in limiting the market power of local generators and facilitating market integration of the NEM \citep{agpc13}, by allowing electricity to be imported from a low priced region with surplus generation capacity to a high priced region. However, the benefits of interregional trades are limited by the physical capacity of interconnectors.
Without sufficient available capacity, the interconnectors during peak-load occasions may become congested, limiting generation capacity within the local market and leading to unnecessarily high prices in the high demand region while the prices in other regions remain at normal levels.
By investigating interconnector usage, \cite{nepal2013testing} find a significant transmission bottleneck for all interconnectors in the NEM and therefore propose that more investment is needed both to increase the capacity of existing interconnectors and to build new interconnection facilities.
The issue of potential under-investment in the market interconnection has also been pointed out by \cite{HiggsH2009}, \cite{garnaut11} and \cite{apergis2016integration}.
As a result, while significant co-movement of spot prices in different regions exists in the NEM, periods of substantial price difference are also observed.
Meanwhile, interregional trade in the NEM still represents a relatively small fraction of the supply for each regional market which is on average below 20\% \citep{aer15}.

Among the five regional markets, QLD and VIC are the main electricity exporters with high capacity of relatively cheap coal generation. NSW, as the largest regional market with the highest average demand in the NEM, typically imports electricity from QLD and VIC to supply around 10\% of its regional demand. With a large share of electricity generation from intermittent resources (i.e. wind), SA has significantly more volatile market supply and prices compared to other regions. As a result, it often imports electricity from VIC for 10\% to 20\% of its local demand, but can also export electricity when sufficient wind power is in place \citep{aer15}. Similarly, TAS, with close to 90\% hydro power in its total registered capacity, has a trade position fluctuating corresponding to relevant market policies\footnote{
	For example, when the Carbon Pricing Mechanism which favours renewable generation was in place between 1 July 2012 and 30 June 2014, TAS became the major electricity exporter and even recorded the highest export ratio of all regions since the NEM started.
} and weather conditions.

\section{The Extremogram}\label{extremo}
Extreme events, by definition, refer to observations in the (left or right) tail of a distribution that occur infrequently by nature. In their own right, the study of extreme events is of importance. More interestingly, in time series data, temporal dependence between extreme events can usually be observed within a single process or between two or more series. The extremogram \citep{davis2009,davis2011,davis2012} is a flexible tool for quantifying various forms of extremal dependence in stationary time series. The extremogram computes the impact that a large value of the time series, or extreme value, has on a future value of the same time series or another time series, $h$ time lags ahead.
Since electricity spot prices are well known for their heavy tails (i.e. spikes) and clustering of high volatility and extreme events \citep{Weron2006,HiggsH2009}, the application of the extremogram can provide important insights in the context of electricity markets.

\subsection{The univariate extremogram}
\cite{davis2009} and \cite{davis2011,davis2012} define the univariate extremogram for a stationary time series ($X_t$) and two sets $A$ and $B$ which are bounded away from zero\footnote{
	A set $S$ is bounded away from zero if a positive $r$ exists such that $S \subset \{y: |y| > r\}$\citep{davis2009,davis2011,davis2012}.
} as
\begin{equation}\label{eq:ext_eq0}
\rho_{A,B}(h) = \lim\limits_{x \rightarrow \infty} P(x^{-1}X_h \in{B}| x^{-1}X_0 \in{A}),
\end{equation}
for $h$ lags in time. Given that $A$ and $B$ are bounded away from zero, the events $\{x^{-1}X_0 \in A\}$ and $\{x^{-1}X_0 \in B\}$ are extreme as the probabilities converge to zero when $x \rightarrow \infty$.

Empirically, the limit on $x$ in \Cref{eq:ext_eq0} can be replaced by a high quantile of the time series. Assuming $a_m$ is the ($1-\frac{1}{m}$)-quantile of the stationary distribution of a time series $\|X_t\|$, the empirical univariate extremogram with $h$ time lags (i.e. $h$ periods ahead), based on the observations $X_{1},...,X_{n}$ is given by
\begin{equation}\label{eq:ext_eq1}
\hat{\rho}_{A,B}(h) = \frac{\sum_{t=1}^{n-h}I\{a_{m}^{-1}X_{t+h}\in{B},a_{m}^{-1}X_{t}\in{A}\}}{\sum_{t=1}^{n}I\{a_{m}^{-1}X_{t}\in{A}\}}.
\end{equation}
In practice, $a_m$ is often replaced by an empirical quantile of the sample $\|X_t\|$.
For example, if we are interested in large positive extreme events (or the right tail), we could use the upper empirical quantiles of the time series for $a_m$ (e.g., the $90^{th}$, $95^{th}$ and $99^{th}$ quantile). Large negative extreme events (or the left tail) can also be examined by the lower empirical quantiles of the time series for $a_m$ (e.g., the $10^{th}$, $5^{th}$ and $1^{st}$ quantile).

Not only can the extremogram provide evidence of extremal dependence in and between time series, it can also be used to help with the statistical modelling of the time series. For example, the estimation of the sample extremogram for a given time series specifies the extent and patterns of volatility clustering for the time series. Connecting this phenomenon to the properties of prominent time series models such as the GARCH or stochastic volatility models can be informative for model selection.
In time series analysis, models are usually selected based on their fit to the centre of the distribution.  However, they may fit poorly to the (right and left) tails of the distribution.  Therefore, with the aim of modelling the extremes of the distribution, a separate examination of extremograms of the time series would be beneficial.
Furthermore, the quality of the model fit could also be assessed by estimating the sample extremogram of the residuals, and the applied model can be improved and updated based on these results.

\subsection{The cross-extremogram} \label{subsec:bi_ext}
The formula for the extremogram in \Cref{eq:ext_eq1} provides a description of the extremal dependence within a time series.  However, in order to observe the extremal dependence between two or more time series, we consider the cross-extremogram.  For the bivariate time series $(X_{t},Y_{t})_{t\in \mathbb{Z}}$, the cross-extremogram with $h$ lags is defined as
\begin{equation}\label{eq:ext_eq2}
\hat{\rho}_{A,B}(h) = \frac{\sum_{t=1}^{n-h}I\{a_{m,Y}^{-1}Y_{t+h}\in{B},a_{m,X}^{-1}X_{t}\in{A}\}}{\sum_{t=1}^{n}I\{a_{m,X}^{-1}X_{t}\in{A}\}}.
\end{equation}
\noindent where $A, B$ are sets bounded away from zero; $a_{m,X}$ and $a_{m,Y}$ are replaced by the respective empirical quantiles computed from the time series $(X_{t})_{t=1,...,n}$ and $(Y_{t})_{t=1,...,n}$. The sample cross-extremogram depends on two empirical quantiles ($a_{m,X}$ and $a_{m,Y}$) which are not required to be equal and are subject to the research question to be analysed.

\subsection{Permutation procedure} \label{subsec:perm_pro}
In order to test for the significance of the serial extremal dependence at each lag of the extremogram, we create confidence bounds for the estimated extremogram through a random permutation procedure.  By permuting the data in the time series, the serial dependence inherent in the data is completely destroyed.  Hence, under the null hypothesis of no serial extremal dependence, we can control the type I error for significance of the extremogram.  The confidence bounds created by the permutation procedure can be considered as the analogue of the confidence bounds calculated as $\pm\frac{z_{1-\frac{\alpha}{2}}}{\sqrt{n}}$ (where $alpha$ is the determined significance interval, $n$ is the number of observations, and $z$ is the cumulative distribution function of the standard normal distribution) which is widely used in the sample autocorrelation function (ACF) of a time series. Spikes that extend beyond the bounds infer that the ACF at the corresponding lags are significantly non-zero and there is serial autocorrelation in the data \citep{little2013oxford}.

In particular, the permutation procedure for the sample univariate extremogram consists of generating a pseudo time series $(X^{*}_{t})_{t\in \mathbb{Z}}$ by permuting the original time series $(X_{t})_{t\in \mathbb{Z}}$. The sample extremogram is then estimated on the pseudo time series using \Cref{eq:ext_eq1}. For the cross-extremogram, bivariate pseudo time series $(X_{t}^{*},Y_{t}^{*})_{t\in \mathbb{Z}}$ are generated and the cross-extremogram is then estimated on the bivariate pseudo time series using \Cref{eq:ext_eq2}.
The result for each lag is recorded and this procedure is repeated a large number of times (in this study, the procedure is repeated 1,000 times).  At each lag of the extremogram, we combine the results across all permutations to create a permutation distribution for the sample extremogram $\hat{\rho}_{A,B}(h)$. We then calculate the (1-$\alpha$/2) and $\alpha$/2 empirical quantiles of the permutation distribution at each lag and interpret them as 100(1-$\alpha$)\% confidence bounds. For each lag $h$, if the estimated sample extremogram $\hat{\rho}_{A,B}(h)$ exceeds the 100(1-$\alpha$)\% confidence bounds obtained from the permutation procedure, we conclude that there is significant evidence of extremal dependence.  On the other hand, if the estimated extremogram $\hat{\rho}_{A,B}(h)$ is within the 100(1-$\alpha$)\% confidence interval, we conclude that there is no significant extremal dependence at that lag.
Since the numerator of the extremogram calculation in \Cref{eq:ext_eq1,eq:ext_eq2} is a sum over $n-h$ terms, the estimated extremogram is dependent on the selection of $h$, however the dependence is only mild due to the random permutation procedure and the infrequent nature of extreme events.
Apart from this boundary effect, the permutation distribution of the sample extremogram is almost the same across all lags $h$. Therefore, in the following graphical assessment of extremograms in \Cref{empirical}, we examine the significance of the estimated extremogram against the (1-$\alpha$/2) and $\alpha$/2 empirical quantiles of the permutation distribution at lag $1$, for interpretational and visual simplicity.

\section{Empirical analysis}\label{empirical}

\subsection{The data}

\Cref{table:descriptive data-entire sample} provides descriptive statistics for the high frequency spot prices in the five regional markets in the NEM: NSW, QLD, SA, TAS and VIC.
In our empirical analysis, both half-hourly trading prices (Panel (a)) and 5-minute dispatch prices (Panel (b)) are used with a sample period from 1 July 2009 to 30 June 2018.

We find that for both half-hourly and 5-minute data, mean price levels are the lowest in VIC, where average prices are below A\$50/MWh, while slightly higher mean price levels could be observed for NSW (A\$52.02/MWh) and TAS (A\$53.78/MWh). Prices throughout the sample period are the highest in SA with a mean price level of A\$61.92/MWh. Note that prices in SA also have the highest standard deviation, indicating the highest volatility of the SA market, while the regional market in TAS has the lowest price standard deviation during the sample period. The data also show typical features for electricity spot prices, like extreme maximum and negative minimum values\footnote{Negative price can be commonly observed in electricity markets \citep{fanone2013case}. Because of the non-storable nature of electricity, it typically happens when electricity demand is low and generators in the market are unable to adjust their output accordingly due to high costs and inflexibility of power plants to shut down and restart.}, high standard deviation, skewness and excess kurtosis, with prices being right-skewed such that the median is always significantly lower than the mean.

\Cref{table:descriptive data-entire sample} also reports the number and percentage of extreme observations where prices are above A\$150, A\$300 and A\$5000 per MWh.
It is noteworthy that QLD and SA appear to be more spiky than other markets, since many more observations of extreme prices are found in these two markets, when the threshold is set as A\$300 or A\$5000 per MWh. In comparison, more smaller price spikes are observed in TAS when the threshold is set as A\$150. Furthermore, the augmented Dickey-Fuller (ADF) test \citep{dickey1979distribution} statistics suggest that all series are stationary at the 1\% significance level.\footnote{Note that typically the literature on modelling and analysing electricity spot prices suggests representing the spot or system price $P_t$ by a sum of two independent parts: a predictable seasonal component $f_t$ and a stochastic component $X_t$, i.e., $P_t = f_t + X_t$, see, e.g., \cite{Weron2006} and \cite{BieMennRachevTrueck}. Therefore, in many empirical applications the first step is to estimate the seasonal component that is assumed to be deterministic, while in a second step the behaviour of the stochastic price component is modelled. While Australian electricity prices also exhibit some degree of seasonality \citep{HiggsH2009,IgnTr:2016}, our analysis is mainly concerned with extreme price spikes above a relatively high threshold. For these normally short-lived and generally unanticipated extreme changes in the spot price, the seasonal pattern only plays a minor role. Thus, we believe that for our analysis of autocorrelation in extreme events as well as the persistence and joint occurrence of extreme price observations with the extremogram, it is appropriate to use actual spot prices instead of a deseasonalised version of these prices.}

\begin{table}[!htbp]
\centering
\footnotesize
\caption[Descriptive statistics]{Descriptive statistics for half-hourly and 5-minute spot prices (A\$/MWh) for the NSW, QLD, SA, TAS and VIC electricity markets over the entire sample period (1 July 2009 to 30 June 2018). }
\label{table:descriptive data-entire sample}
\begin{tabular*}{1\textwidth}{@{\extracolsep{\fill}}rrrrrr}
  \hline
 & NSW & QLD & SA & TAS & VIC \\
  \hline
  \multicolumn{6}{l}{\textit{Panel (a) Half-Hourly Spot Prices (A\$/MWh), 157,776 obs}} \\
  &&&&&\\
Mean & 52.02 & 55.25 & 61.92 & 53.78 & 48.32 \\
Med. &     41.78 &     40.62 &     42.25 &     39.33 &     37.89 \\
  Max. & 14000.00 & 13882.77 & 14166.50 & 12400.26 & 12931.04 \\
  Min. & -264.31 & -1000.00 & -996.70 & -999.64 & -817.03 \\
  Std. dev. & 163.05 & 200.49 & 251.39 & 97.24 & 132.84 \\
  Skew. & 51.38 & 36.91 & 33.91 & 66.30 & 56.83 \\
 Kurt. & 3186.60 & 1819.94 & 1383.26 & 7344.94 & 3901.76 \\
 Adf   & -94.70*** &    -93.47*** &    -95.30*** &    -67.59*** &    -93.06*** \\
Spike obs. (\textgreater A\$150) &2145 & 3008 & 4832 & 5089 & 2136 \\
  Percentage & 1.36\% & 1.91\% & 3.06\% & 3.23\% & 1.35\% \\
Spike obs. (\textgreater A\$300) & 227 & 713 & 1044 & 426 & 188 \\
  Percentage & 0.14\% & 0.45\% & 0.66\% & 0.27\% & 0.12\% \\
Spike obs. (\textgreater A\$5000) & 47 & 48 & 84 & 7 & 32 \\
Percentage & 0.03\% & 0.03\% & 0.05\% & 0.00\% & 0.02\% \\

		&&&&&\\
		\multicolumn{6}{l}{\textit{Panel (b) 5-minute Spot Prices (A\$/MWh), 946,656 obs}} \\
		&&&&&\\
		Mean & 52.02 & 55.25 & 61.92 & 53.78 & 48.32 \\
		Med.&     41.83 &     40.05 &     42.05 &     39.26 &     37.80 \\
		Max. & 14200.00 & 14200.00 & 14200.00 & 14200.00 & 13800.00 \\
		Min. & -1000.00 & -1000.00 & -1000.00 & -1000.00 & -1000.00 \\
		Std. dev. & 196.97 & 305.29 & 338.55 & 139.54 & 169.65 \\
		Skew. & 54.23 & 38.82 & 33.94 & 73.64 & 60.46 \\
		Kurt. & 3200.53 & 1591.59 & 1208.29 & 6362.72 & 3951.53 \\
		Adf &   -147.43*** &   -145.74*** &   -139.26*** &   -176.19*** &   -154.61*** \\
		Spike obs. (\textgreater A\$150) &11524 & 15877 & 26081 & 30631 & 12377 \\
		Percentage & 1.22\% & 1.68\% & 2.76\% & 3.24\% & 1.31\% \\
		Spike obs. (\textgreater A\$300) & 1138 & 2752 & 6291 & 3171 & 1206 \\
		Percentage & 0.12\% & 0.29\% & 0.66\% & 0.33\% & 0.13\% \\
		Spike obs. (\textgreater A\$5000) & 326 & 646 & 806 & 122 & 236 \\
		Percentage & 0.03\% & 0.07\% & 0.09\% & 0.01\% & 0.02\% \\
		\hline
	\end{tabular*}
\begin{flushleft}
	\footnotesize{\emph{Notes}: The hypotheses of the ADF test are $H_0$: a unit root (non-stationary); $H_1$: no unit root (stationary). The null hypothesis is rejected under a particular significance level if the test statistic is below the critical values which are -2.57 (10\%), -2.86 (5\%) and -3.44 (1\%).}%
\end{flushleft}

\end{table}

\begin{figure}[!h]

\centering
\includegraphics[width=1\textwidth]{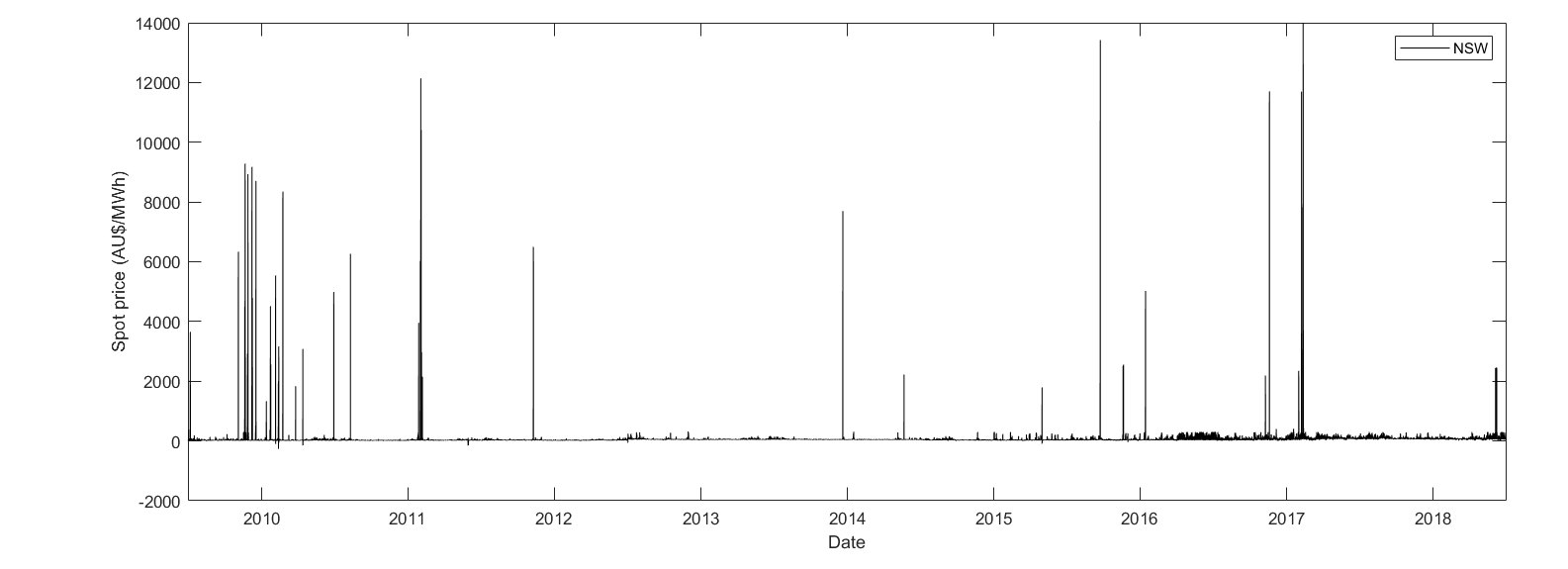}
\includegraphics[width=1\textwidth]{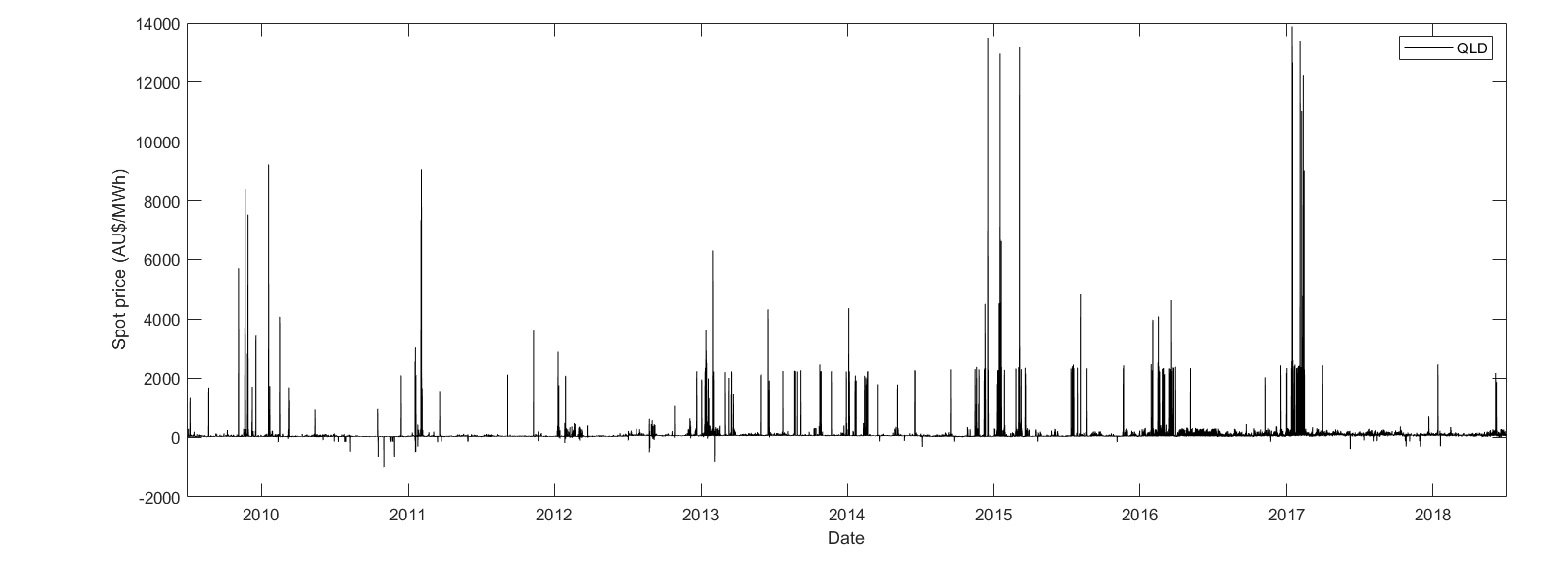}
\caption[Half-hourly spot price plots]{Half-hourly spot prices (A\$/MWh) for the NSW (upper panel)and QLD (lower panel) electricity markets from 1 July 2009 to 30 June 2018.}
\label{fig:price plot}
\end{figure}

\Cref{fig:price plot} plots the spot prices over the entire sample period for two regions: NSW (upper panel) and QLD (lower panel). Significant price spikes are observed in the figure. Within each regional market, clusters of spikes are evident. When considering both regions, there are several occasions when spikes jointly appear in NSW and QLD. Time series plots for other regions are presented \Cref{Afig:price plot} in the Appendix.

Note that our analysis focuses on extreme outcomes only in the upper tail region. While prices in the NEM very rarely also exhibit negative price spikes, the major interest of market participants is typically extreme positive price observations. The magnitude of negative price spikes is much smaller than that of positive ones, because the price floor of the NEM is (-A\$1000)/MWh, whereas the cap is up to A\$14,200/MWh during our sample period \citep{AEMO2014}.
With the occurrence of extremely high prices, profits of market participants in the NEM can be wiped out or significantly reduced within a short period of time. Both the size and the duration of these extreme price spikes are of substantial concern, particularly to retailers who buy electricity for a flexible price but sell it for a fixed price \citep{eichler2014models,clements2015modelling,AEMO2014}.

\subsection{Univariate analysis}

\subsubsection{Analysis of half-hourly prices}

We begin our examination with a univariate analysis of half-hourly prices, i.e. we compute the extremograms for each individual market in the NEM separately.\footnote{Both the univariate and cross-extremograms were estimated using the extremogram R package \citep{frolova}.} Such an analysis will provide us with a better understanding of the persistence of extreme price outcomes.

As pointed out by \cite{davis2009} and \cite{davis2011,davis2012}, one of the key issues in computing the extremogram is the selection of a suitable threshold to define `extreme events'. As mentioned earlier, ASX Energy offers `\$300 Cap Products', i.e. a derivatives contract specifically designed for the occurrence of extreme price observations. The payoff of these products is determined by the sum of all half-hourly spot prices greater than A\$300/MWh for a calendar quarter in a regional market and is highly relevant to our research objective in terms of assessing the pattern of extreme price events. Based on the definition of these products that are widely traded in the market, we believe that A\$300/MWh is a natural choice for an appropriate threshold, above which the prices are considered as `extreme'. Note that this threshold corresponds to the $99.86^{th}$ percentile for NSW, the $99.55^{th}$ percentile for QLD, the $99.34^{th}$ percentile for SA, the $99.73^{th}$ percentile for TAS and the $99.88^{th}$ percentile for VIC such that less than 1\% of half-hourly spot electricity prices are above A\$300/MWh.

\begin{figure}[!h]
	\centering
	\begin{minipage}[t]{1\linewidth}	
		\includegraphics[width = 1\linewidth]{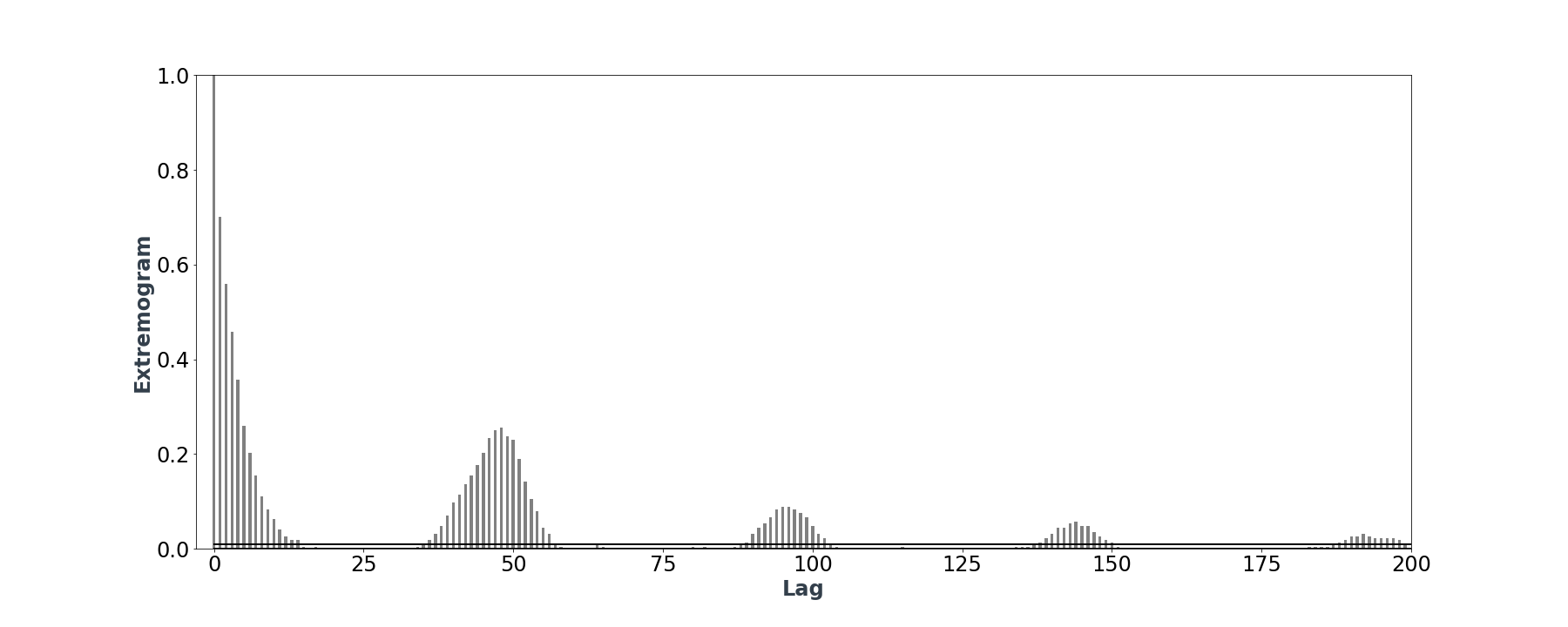}
		\vspace{-2em}
		\subcaption*{\footnotesize{A: NSW}}
	\end{minipage}
	\begin{minipage}[t]{1\linewidth}	
		\includegraphics[width = 1\linewidth]{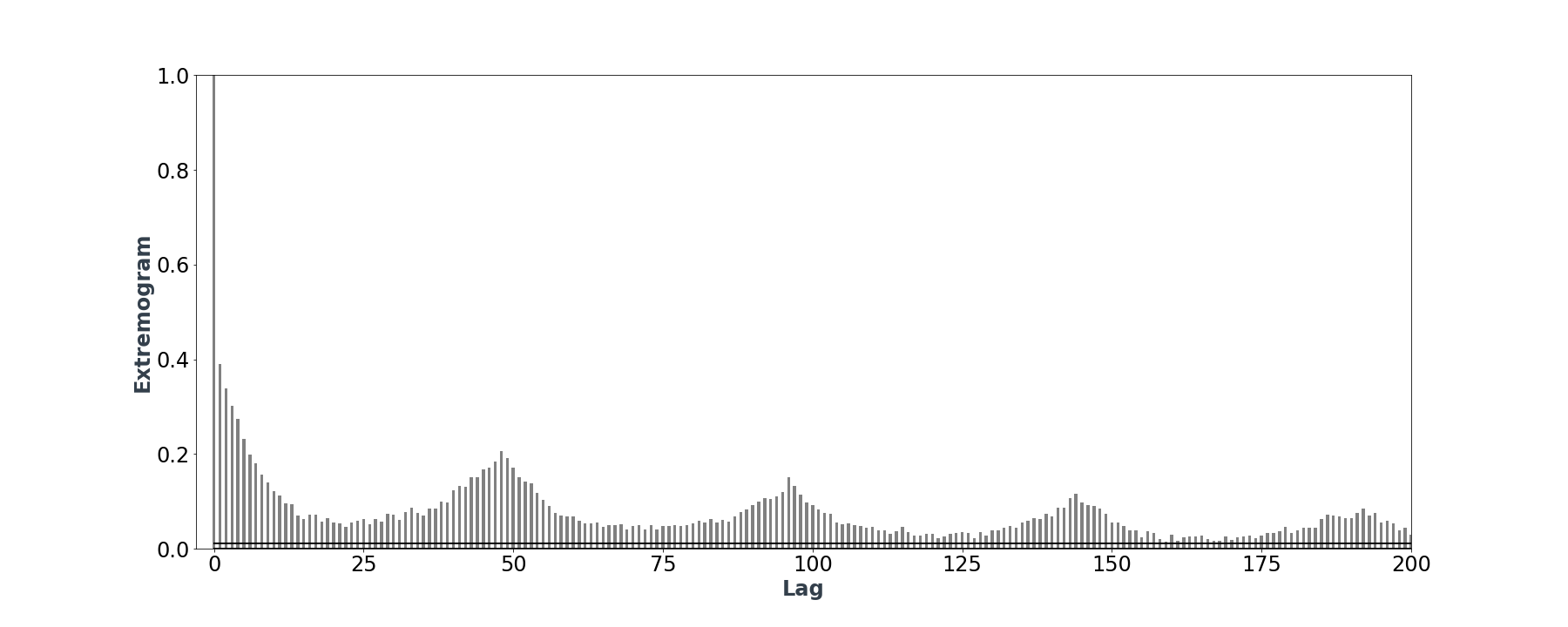}
		\vspace{-2em}
		\subcaption*{\footnotesize{B: QLD}}
	\end{minipage}
	\begin{minipage}[t]{1\linewidth}	
		\includegraphics[width = 1\linewidth]{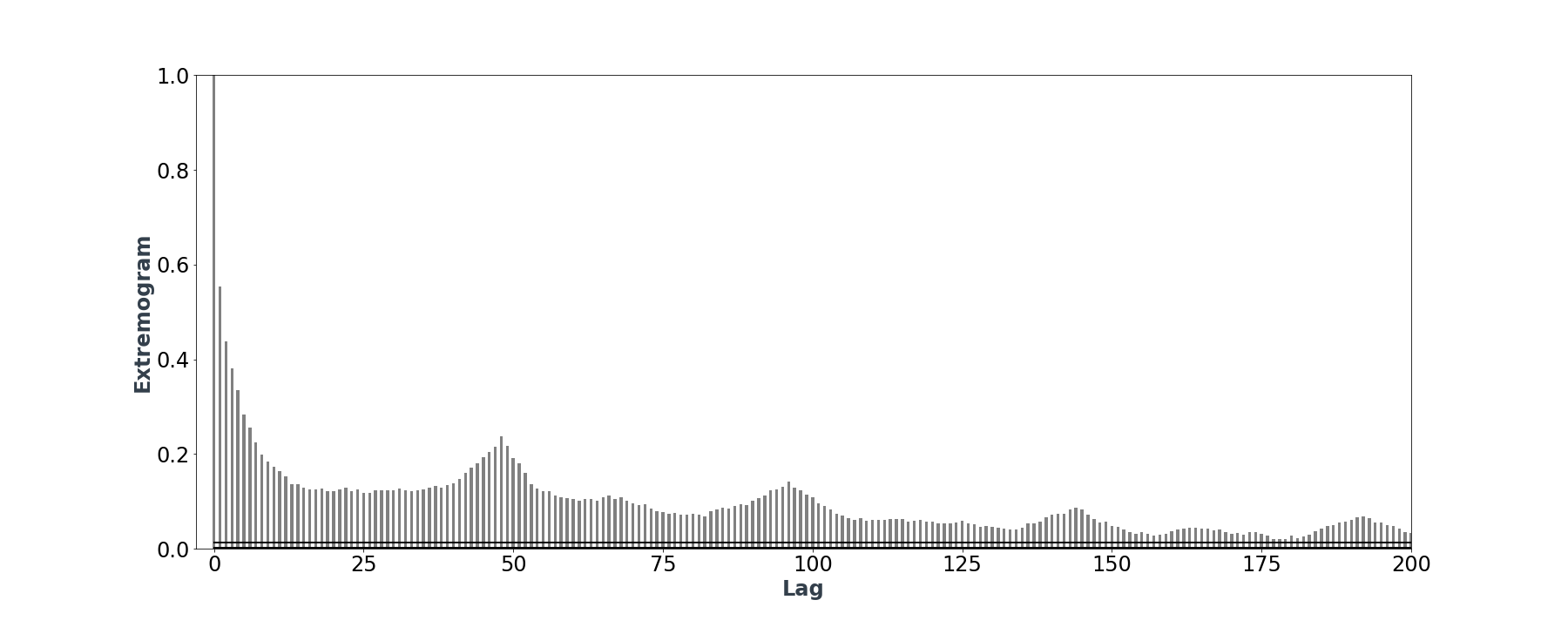}
		\vspace{-2em}
		\subcaption*{\footnotesize{C: SA}}
	\end{minipage}
	\caption[Univariate extremograms (half-hourly) in NSW, QLD and SA]{The empirical univariate extremogram of half-hourly spot electricity prices in (A) NSW, (B) QLD, and (C) SA, based on a sample period from 1 July 2009 to 30 June 2018.}
	\label{fig:uni_ext1}
\end{figure}

\Cref{fig:uni_ext1} provides a plot of the sample extremograms applied to half-hourly price observations for the markets in NSW, QLD and SA for the considered sample period. In each graph, the solid horizontal line is the generated 99\% confidence bound, based on the permutation procedure described above.
Sample extremograms outside these bounds indicate the existence of significant serial extremal dependence at the corresponding lags.

For the NSW market, the extremogram has a large value (about 0.75) at the first lag, indicating that given a price spike in NSW a half hour ago, there is an approximately 75\% probability of another extreme price in this market in the current half hour. The extremal dependence as measured by the extremogram diminishes rapidly, and becomes insignificant after $12$ lags. However, a cluster of significant spikes reappear around the $48^{th}$ lag. Weak but significant extremal dependence is also found around the $96^{th}$ and $144^{th}$ lags.
The sample data are recorded on a half-hourly basis, and each day has $48$ half-hourly intervals.
Therefore, the univariate extremogram plots for NSW illustrate the existence of extremal dependence between prices one day apart, indicating that given a price spike on a certain day, more spikes tend to occur in this market around the same time on the following days.
The extremogram plots for QLD are different in shape compared to those for NSW. Significant extremal dependence is observed at all lags but becomes weaker more gradually. The slow decay of extremograms over long lags indicates strong persistence of extreme prices in the QLD market. Apart from the difference in persistence, compared to the NSW sample extremogram, the QLD sample extremogram exhibits similar periodicity such that the value of extremogram spikes peak every $48$ lags, indicating that a given spike has a higher possibility of reappearing at the same time in the following days.
The extremogram plot for prices in SA show a pattern similar to the QLD market. In particular, the extremogram spikes are persistent, and the persistence of price spikes is even stronger than in QLD. Furthermore, we observe relatively high values for the extremogram around the $1^{st}$, $48^{th}$, $96^{th}$, $144^{th}$ and $192^{th}$ lags.

\begin{figure}[!h]
	\centering
	\begin{minipage}[t]{1\linewidth}	
		\includegraphics[width = 1\linewidth]{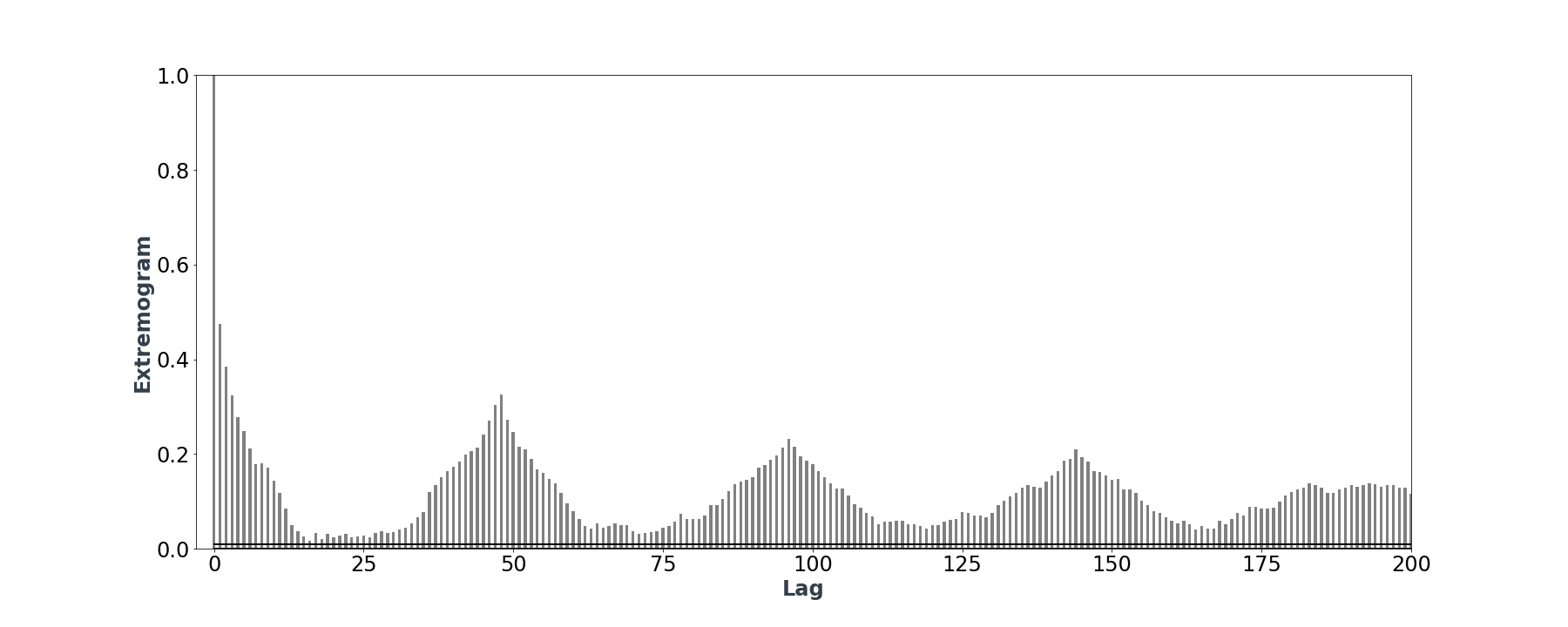}
		\vspace{-2em}
		\subcaption*{\footnotesize{A: TAS}}
	\end{minipage}
	\begin{minipage}[t]{1\linewidth}	
		\includegraphics[width = 1\linewidth]{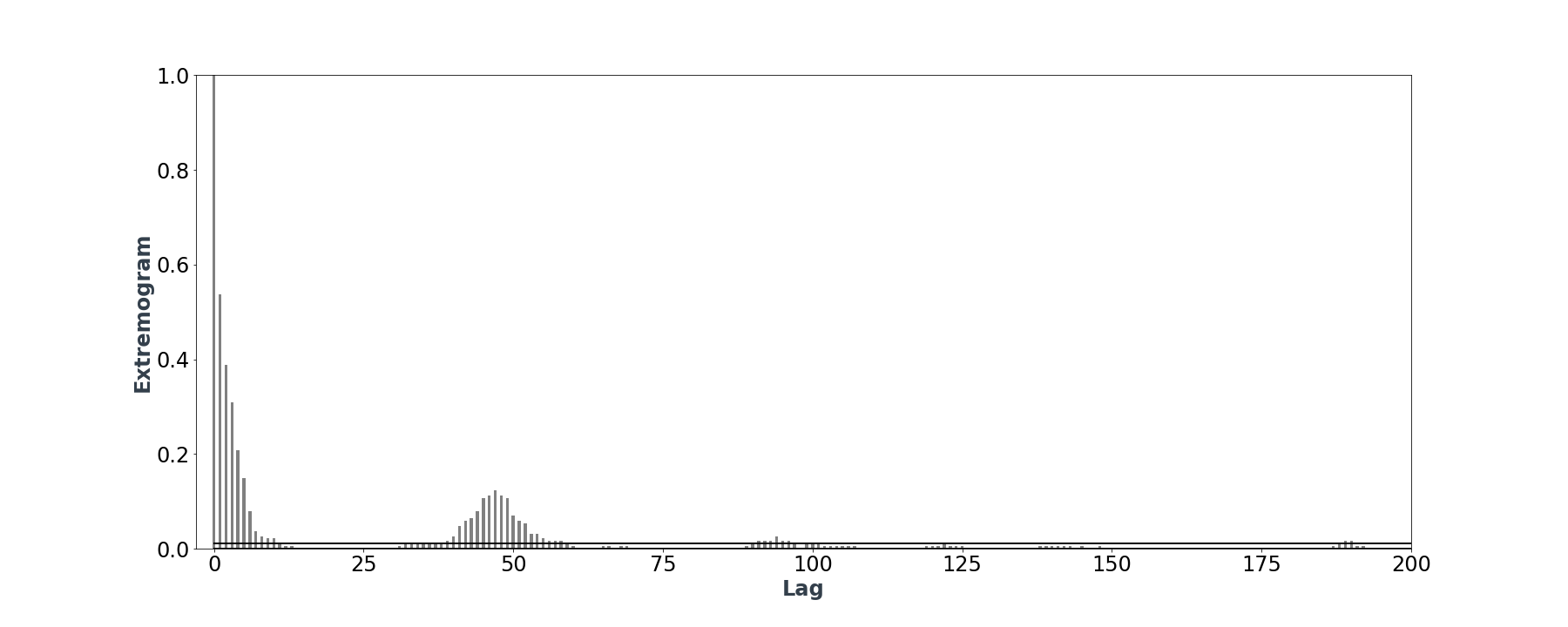}
		\vspace{-2em}
		\subcaption*{\footnotesize{B: VIC}}
	\end{minipage}
	\caption[Univariate extremograms (half-hourly) in TAS and VIC]{The empirical univariate extremogram of half-hourly spot electricity prices in (A) TAS and (B) VIC, based on a sample period from 1 July 2009 to 30 June 2018.}
	\label{fig:uni_ext2}
\end{figure}

\Cref{fig:uni_ext2} plots the sample extremograms applied to half-hourly price observations for the electricity markets in TAS and VIC.
The TAS market shows evidence of relatively strong persistence of extreme prices, since extremogram spikes over the first $200$ lags are all significant with a slow speed of decay.
In particular, the extreme price persistence after the 100$^{th}$ lag in TAS is the strongest among all regional markets in the NEM.
The possibility of extreme prices reappearing around the same time in the following days in TAS is also stronger than for other markets.

On the other hand, for the VIC market, significant extremogram spikes are only observed over the first $14$ lags and around the $48^{th}$ lag. The extremal dependence of prices in VIC decays to zero within $54$ lags, which is the fastest among the five considered markets.

\subsubsection{Analysis of 5-minute prices}\label{section: 5-min empirical}

In this section, 5-minute dispatch data are considered so that more details of extremograms in the high-frequency domain can be examined. In addition, the analysis in a 5-minute context is of importance since the NEM is in transition from 30-minute settlement to 5-minute settlement. In particular, the NEM will start a 5-minute settlement in late 2021 and is argued to be able to provide better price signals for fast response technologies and demand responses\footnote{See https://www.aemc.gov.au/rule-changes/five-minute-settlement for details about 5-minute settlement.}.
With higher frequency of the data, we are able to focus on more severe price spikes because of more available observations of these extreme outcomes. Therefore, an A\$5000/MWh threshold is chosen for 5-minute extremal dependence examination. Also, A\$5000/MWh is a critical threshold since the market regulator is required to prepare a report to investigate the underlying reason causing the high price value whenever this threshold is reached\footnote{For examples of the `\$5000 Report', see https://www.aer.gov.au/taxonomy/term/310.}.

\begin{figure}[!h]
        \centering
        \begin{minipage}[t]{1\linewidth}        
                \includegraphics[width = 0.95\linewidth]{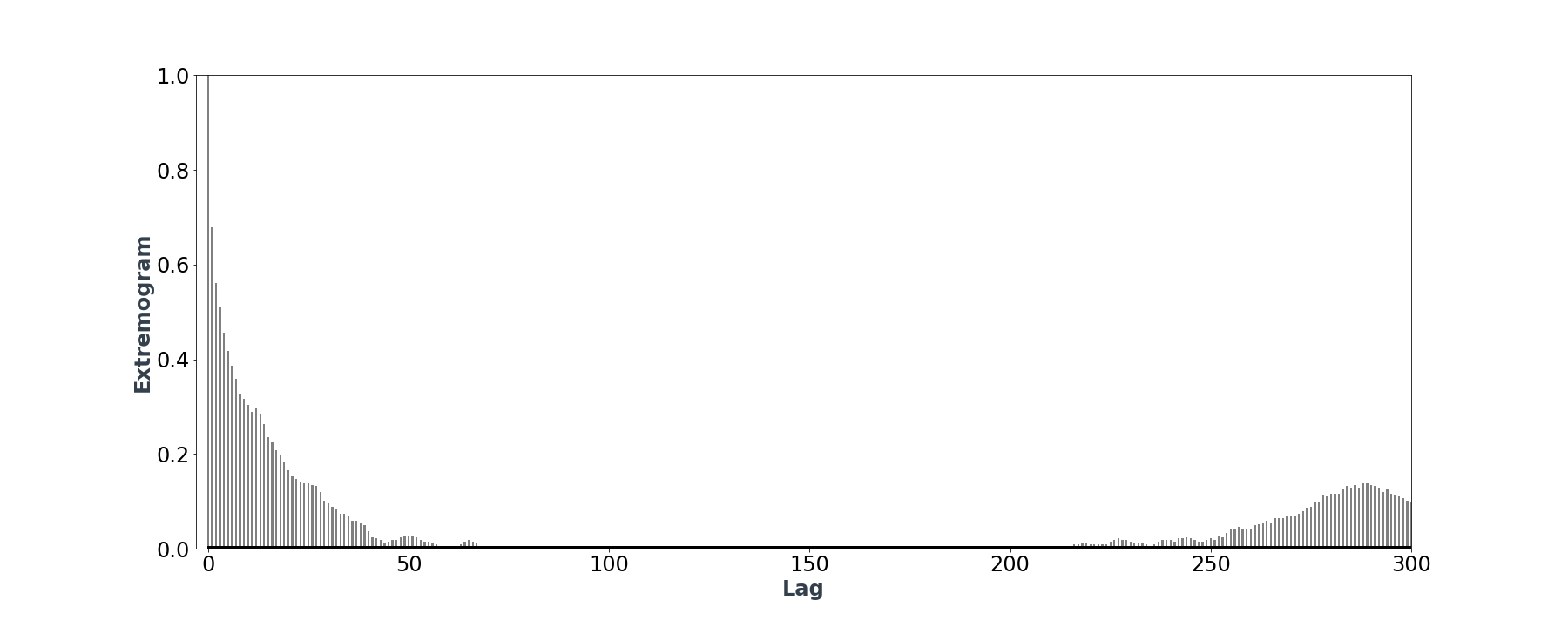}
                \vspace{-2em}
                \subcaption*{\footnotesize{A: NSW}}
        \end{minipage}
        \begin{minipage}[t]{1\linewidth}        
                \includegraphics[width = 0.95\linewidth]{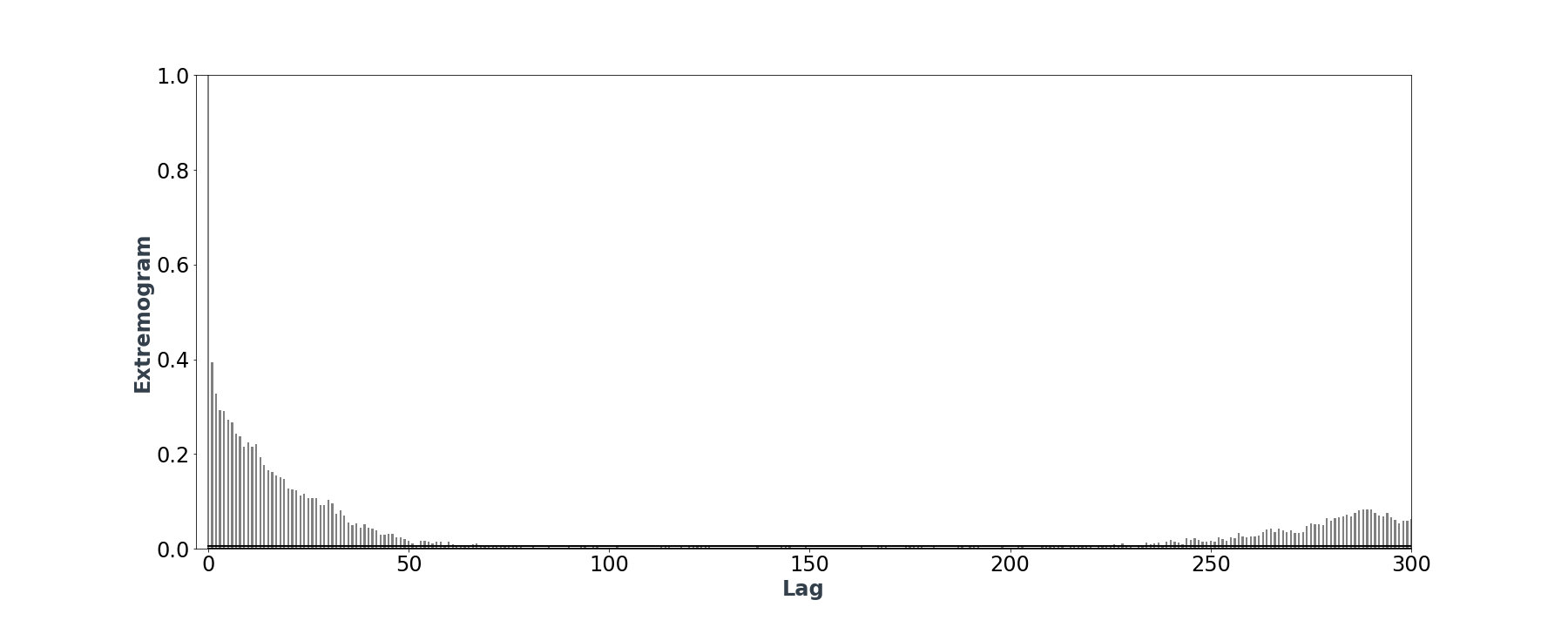}
                \vspace{-2em}
                \subcaption*{\footnotesize{B: QLD}}
        \end{minipage}
        \begin{minipage}[t]{1\linewidth}        
                \includegraphics[width = 0.95\linewidth]{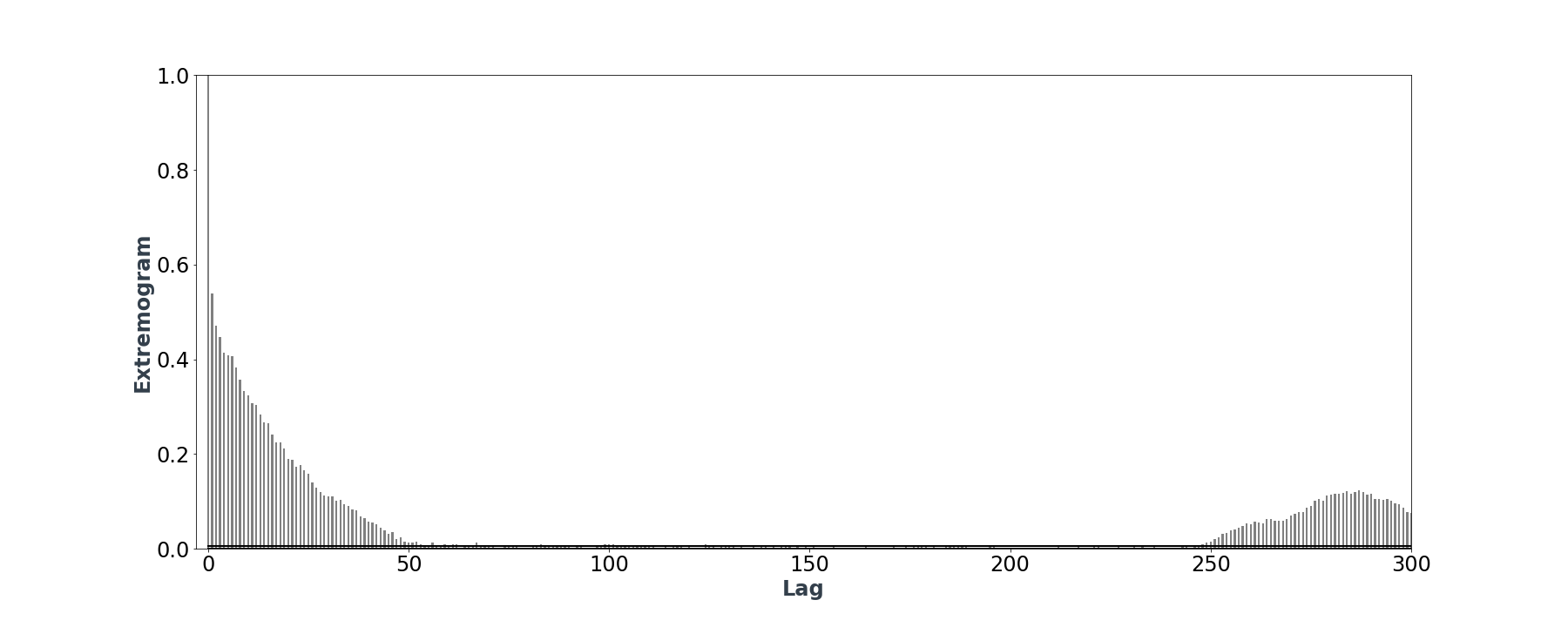}
                \vspace{-2em}
                \subcaption*{\footnotesize{C: SA}}
        \end{minipage}

        \caption[Univariate extremograms (5-minute) in NSW, QLD and SA]{The empirical univariate extremogram of 5-minute spot electricity prices in (A) NSW, (B) QLD, and (C) SA, based on a sample period from 1 July 2009 to 30 June 2018.}
        \label{fig:uni_ext1_5min}
\end{figure}

\begin{figure}[!h]
        \centering
        \begin{minipage}[t]{1\linewidth}        
                \includegraphics[width = 1\linewidth]{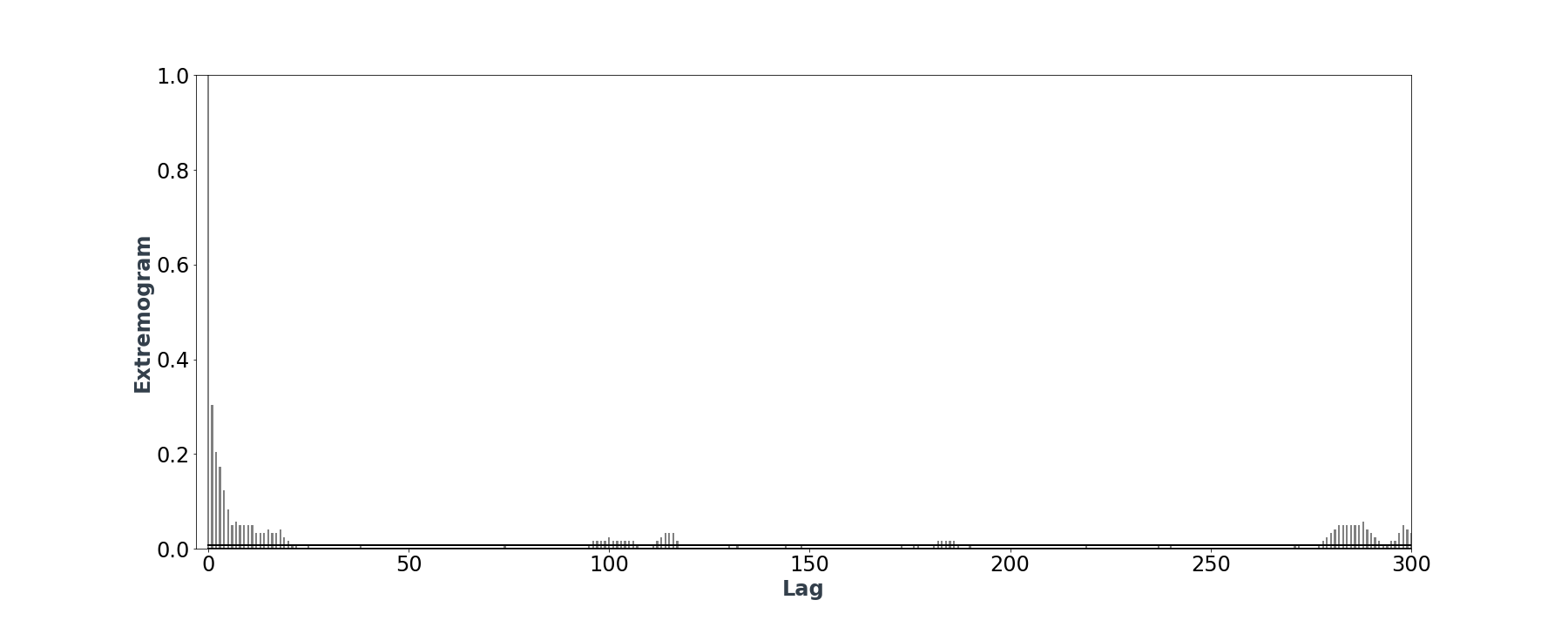}
                \vspace{-2em}
                \subcaption*{\footnotesize{A: TAS}}
        \end{minipage}
        \begin{minipage}[t]{1\linewidth}        
                \includegraphics[width = 1\linewidth]{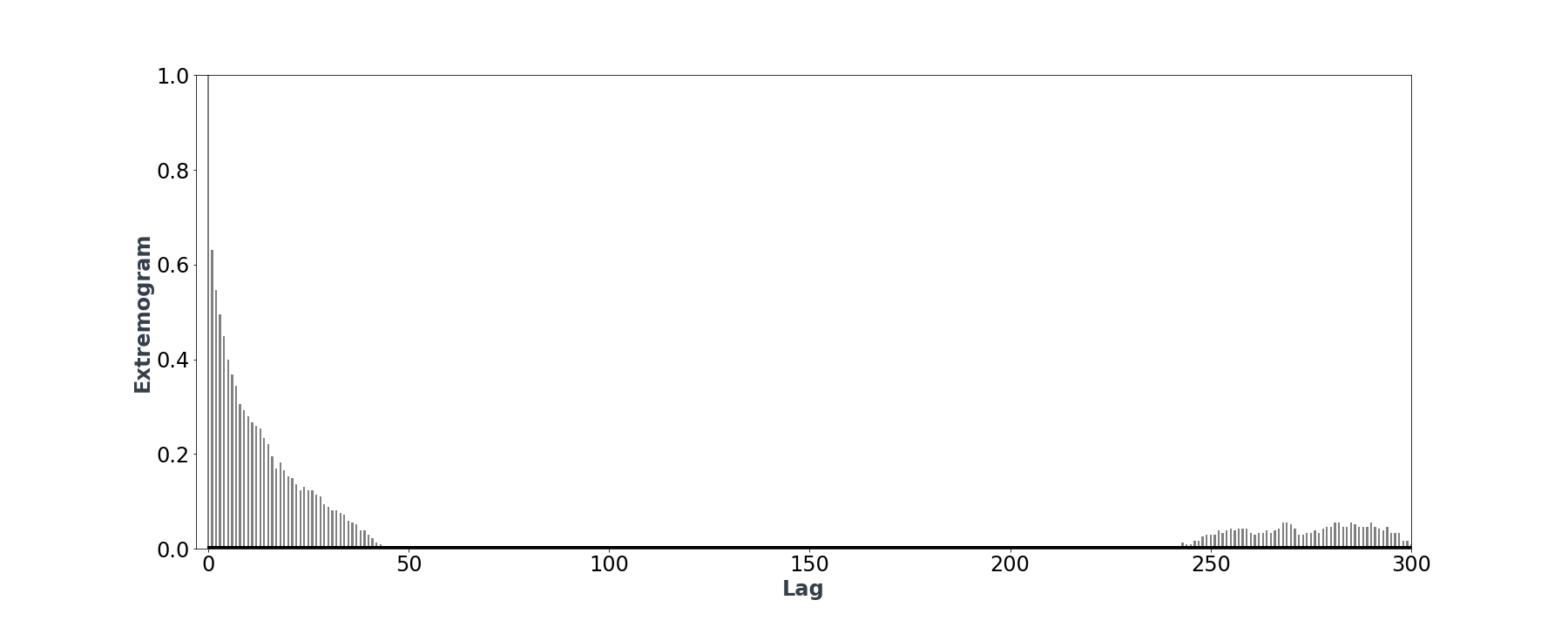}
                \vspace{-2em}
                \subcaption*{\footnotesize{B: VIC}}
        \end{minipage}

        \caption[Univariate extremograms (5-minute) in TAS and VIC]{The univariate empirical extremogram of 5-minute spot electricity prices in (A) TAS and (B) VIC, based on a sample period from 1 July 2009 to 30 June 2018.}
        \label{fig:uni_ext2_5min}
\end{figure}

The estimated 5-minute extremograms are presented in \Cref{fig:uni_ext1_5min} and \Cref{fig:uni_ext2_5min}. The extremograms for all considered markets exhibit significant values following the first lag and around the $288^{th}$ lag.
Since each single day contains $288$ 5-minute intervals, similar to the half-hourly analysis, this means a price spike is more likely to reappear within a series of subsequent intervals and at the same time the next day.
Interestingly, when the A\$5000/MWh threshold is considered, none of the considered markets present strong local persistence of extreme prices, indicating the rarity of events when spot prices exceed A\$5000/MWh.
In between the first lag and around the $288^{th}$ lag, the estimated extremogram spikes decrease to be insignificant typically within $50$ lags (i.e. $4$ hours).
The estimated extremogram spikes are lowest in TAS and decay to zero the fastest, which is consistent with the smallest number of over A\$5000/MWh price observations  for TAS in \Cref{table:descriptive data-entire sample}.

Overall, with regard to univariate extremograms, each of the regional markets in the NEM exhibit persistence of extreme prices to a certain extent.
In addition, for all considered markets, serial extremal dependence tends to be relatively high between prices a day apart, indicating a high possibility of an existing price spike reappearing around the same time after one day.
These features within individual electricity markets are consistent with the findings of \cite{clements2013semi} and \cite{eichler2014models}.
In particular, we find significant persistence in extreme price outcomes in QLD, SA and TAS.
The highest persistence of extreme prices is observed in SA. This is according to our expectations, since SA has the most volatile and spiky spot prices due to its large share of intermittent wind generation. The significant uncertainty in the SA market supply of electricity induces large fluctuations of spot prices. In addition, the risk of interconnectors between SA and VIC being constrained, when SA local demand soars and insufficient wind power is available, contributes to more significant and persistent price spikes.
The high persistence of extreme prices in QLD could be attributed to the high concentration level in this market (i.e. high market power controlled by several market participants) \citep{hurn2016smooth,AEMO2014} which leads to more price spikes.
Similarly, for TAS which is the smallest market in the NEM and distant from mainland regions with only one submarine cable connecting to VIC, the market power is highly concentrated with one government owned participant, i.e. Hydro Tasmania, owning all generation capacity \citep{AEMO2014}. High market concentration as well as the occasional inability to import electricity from VIC when the interconnector is congested could be contributing factors to the relatively high persistence and periodicity of half-hourly prices above the A\$300/MWh threshold in TAS.
Interestingly, however, because of the flexibility of the local hydro power system to rapidly respond to extraordinary market events, very few `huge' spikes (i.e. prices exceeding A\$5000/MWh) are observed in TAS. This leads to a much weaker persistence in the estimated extremograms for the 5-minute analysis for TAS compared to other markets.

We also find that well-connected markets such as NSW and VIC show significantly lower persistence of extreme price outcomes. These findings indicate the significant importance of interconnectors in terms of reducing the occurrence of extreme price outcomes and high volatility periods.
Furthermore,
the 5-minute analysis shows that the persistence existing in prices above A\$5000/MWh is relatively low with prices returning to less extreme levels below the threshold more quickly for all markets. However, similar to the periodicity found in the half-hourly analysis with a A\$300/MWh threshold, a current price spike exceeding A\$5000/MWh also has a chance of reappearing the next day, what is evidenced by low but significant extremograms around the 288$^{th}$ lag.

\subsection{Multivariate analysis}

\subsubsection{Analysis of half-hourly prices}

In this section, we analyse cross-extremal dependence to assess the transmission effects of extreme prices between the regional markets. To do this we compute cross-extremograms in a bivariate setting for different combinations of the regional markets.

\begin{figure}[!htbp]
			\centering
			\includegraphics[width=\linewidth,height=15cm]{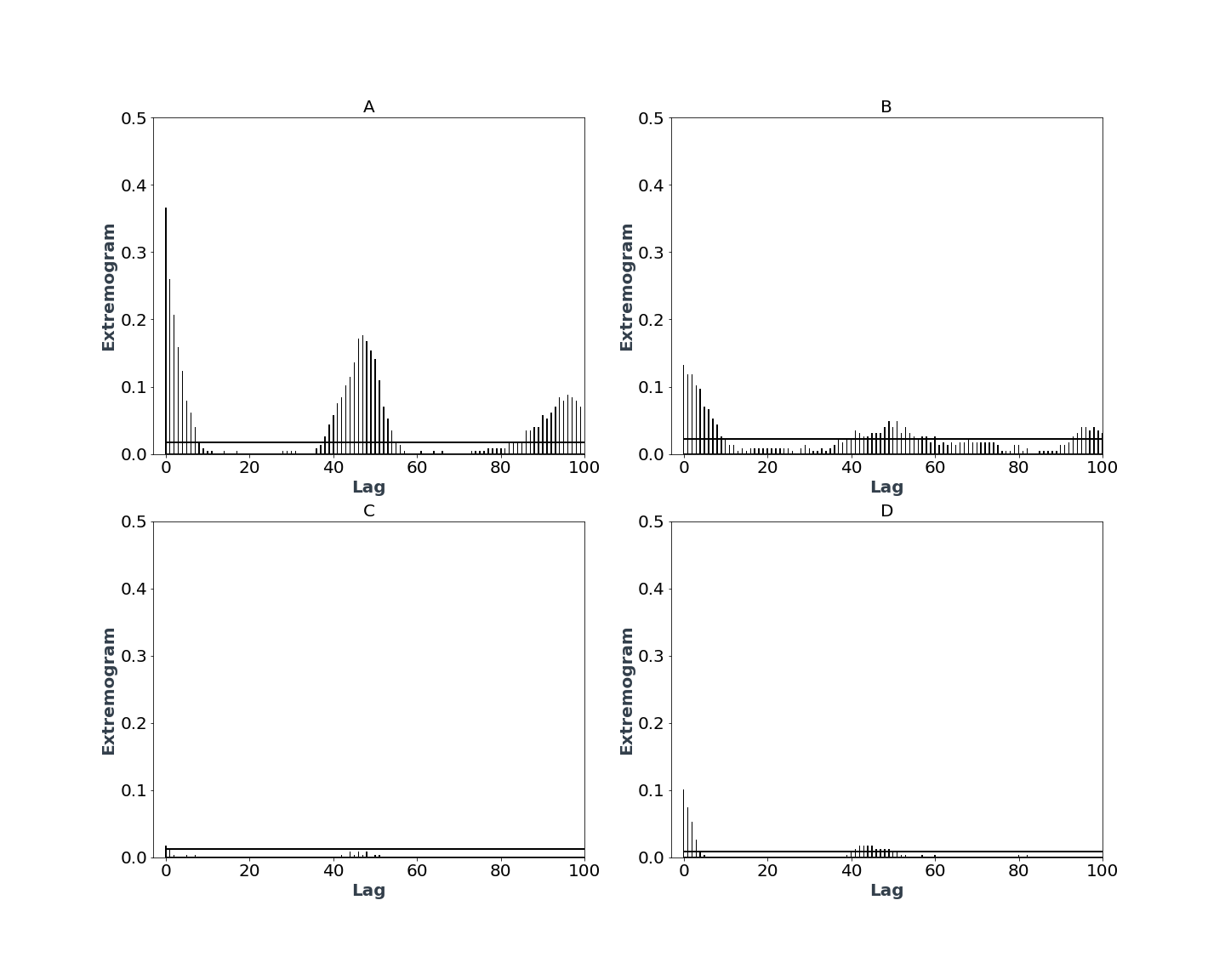}
			\caption[Cross-extremograms (half-hourly) conditioning on price spikes in NSW]{Sample cross-extremograms of half-hourly spot electricity prices conditioning on price spikes in NSW. The plot provides cross-extremograms for (A) NSW $\rightarrow$ QLD, (B) NSW $\rightarrow$ SA, (C) NSW $\rightarrow$ TAS and (D) NSW $\rightarrow$ VIC based on a sample period from 1 July 2009 to 30 June 2018.}
				\label{fig:cross_ext_NSW}
\end{figure}

\Cref{fig:cross_ext_NSW} shows the sample cross-extremograms of half-hourly spot prices conditioning on price spikes in the NSW electricity market. In each panel, the cross-extremogram at lag $0$ can be interpreted as the probability of observing a price spike in the corresponding market given an extreme price observed in NSW at the same time. In comparison, the cross-extremogram at lag $1$ can be interpreted as the probability of observing a price spike in the corresponding market at time $t$ (the current half-hourly interval) given an extreme price observed in NSW at $t-1$ (the previous half-hourly interval). Significant extremal dependence as measured by cross-extremograms is found between NSW and QLD, NSW and SA, and NSW and VIC. In particular, the most significant cross-extremograms are observed between NSW and QLD at lags $0$ to $8$, and around the $48^{th}$ and $96^{th}$ lags which correspond to price spikes at the same time on two consecutive days. The cross-extremograms observed in SA conditioning on spikes in NSW drop off to zero over the first $8$ lags, while the cross-extremograms observed in VIC only last for three lags.

\begin{figure}[!htbp]
			\centering
			\includegraphics[width=\linewidth,height=15cm]{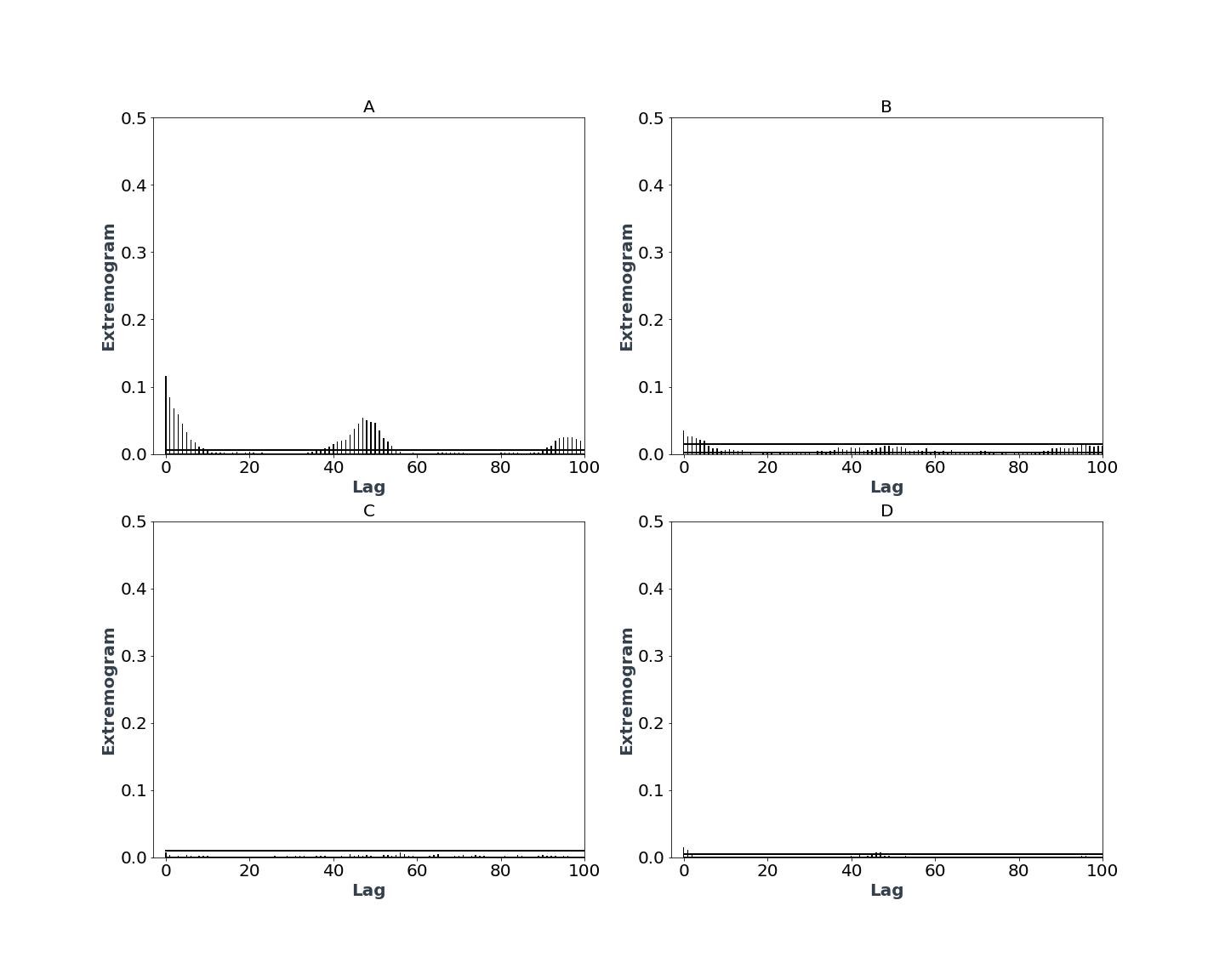}
			\caption[Cross-extremograms (half-hourly) conditioning on price spikes in QLD]{Sample cross-extremograms of half-hourly spot electricity prices conditioning on price spikes in QLD. The plot provides cross-extremograms for (A) QLD $\rightarrow$ NSW, (B) QLD $\rightarrow$ SA, (C) QLD $\rightarrow$ TAS and (D) QLD $\rightarrow$ VIC based on a sample period from 1 July 2009 to 30 June 2018.}
				\label{fig:cross_ext_QLD}
\end{figure}

With regard to cross-extremograms conditioning on price spikes in QLD (\Cref{fig:cross_ext_QLD}), extremal dependence is only observed between QLD and NSW. The significant extremal dependence at lag $0$ indicates the probability of joint price spikes in these two markets. Significant extremal dependence is also found around the $48^{th}$ and $96^{th}$ lags.

\begin{figure}[!htbp]
			\centering
			\includegraphics[width=\linewidth,height=15cm]{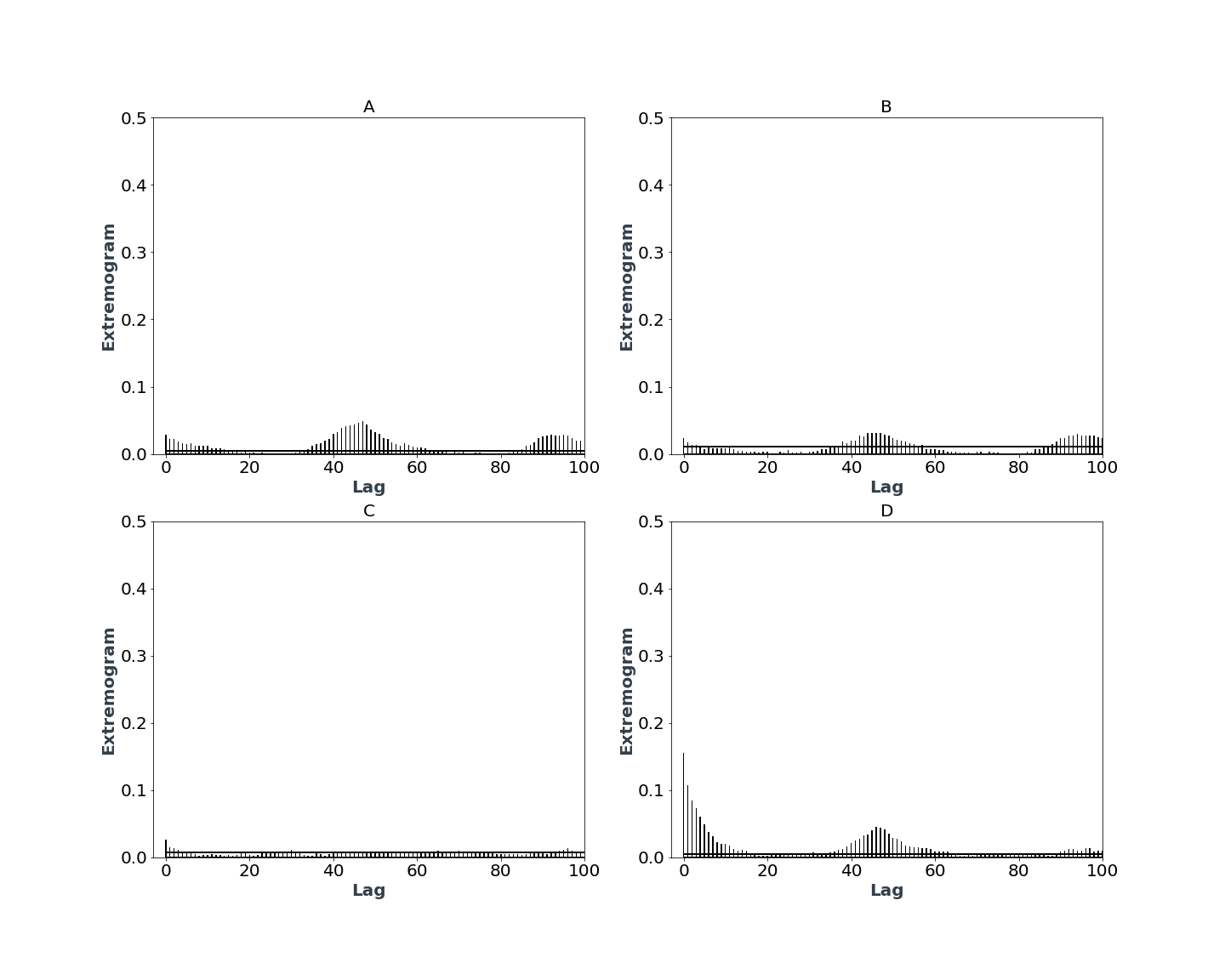}
			\caption[Cross-extremograms (half-hourly) conditioning on price spikes in SA]{Sample cross-extremograms of half-hourly spot electricity prices conditioning on price spikes in SA. The plot provides cross-extremograms for (A) SA $\rightarrow$ NSW, (B) SA $\rightarrow$ QLD, (C) SA $\rightarrow$ TAS and (D) SA $\rightarrow$ VIC based on a sample period from 1 July 2009 to 30 June 2018.}
				\label{fig:cross_ext_SA}
\end{figure}

In \Cref{fig:cross_ext_SA}, significant cross-extremograms are found between SA and all other markets. Significant extremal dependence tends to appear in clusters: over short lags following lag $0$ and around the $48^{th}$ and $96^{th}$ lags. The highest level of extremal dependence is observed between SA and VIC, indicating both contemporaneous and lagged transmission effects of price spikes from SA to VIC. Interestingly, for NSW and QLD, cross-extremograms conditioning on price spikes in SA are larger around the $48^{th}$ and $96^{th}$ lags than at the first several lags. This indicates that extreme prices in SA tend to spill over to NSW and QLD one day and two days later, rather than simultaneously occur in these markets. One reason for this could be the market expectations in NSW and QLD about high demand and price spikes in SA re-occurring around the same hours on the following days, which can lead to changed bidding behaviours in these markets for those hours. Recall that SA is not directly connected to either NSW or QLD, but only indirectly through interconnectors via VIC.

\begin{figure}[!htbp]
			\centering
			\includegraphics[width=\linewidth,height=15cm]{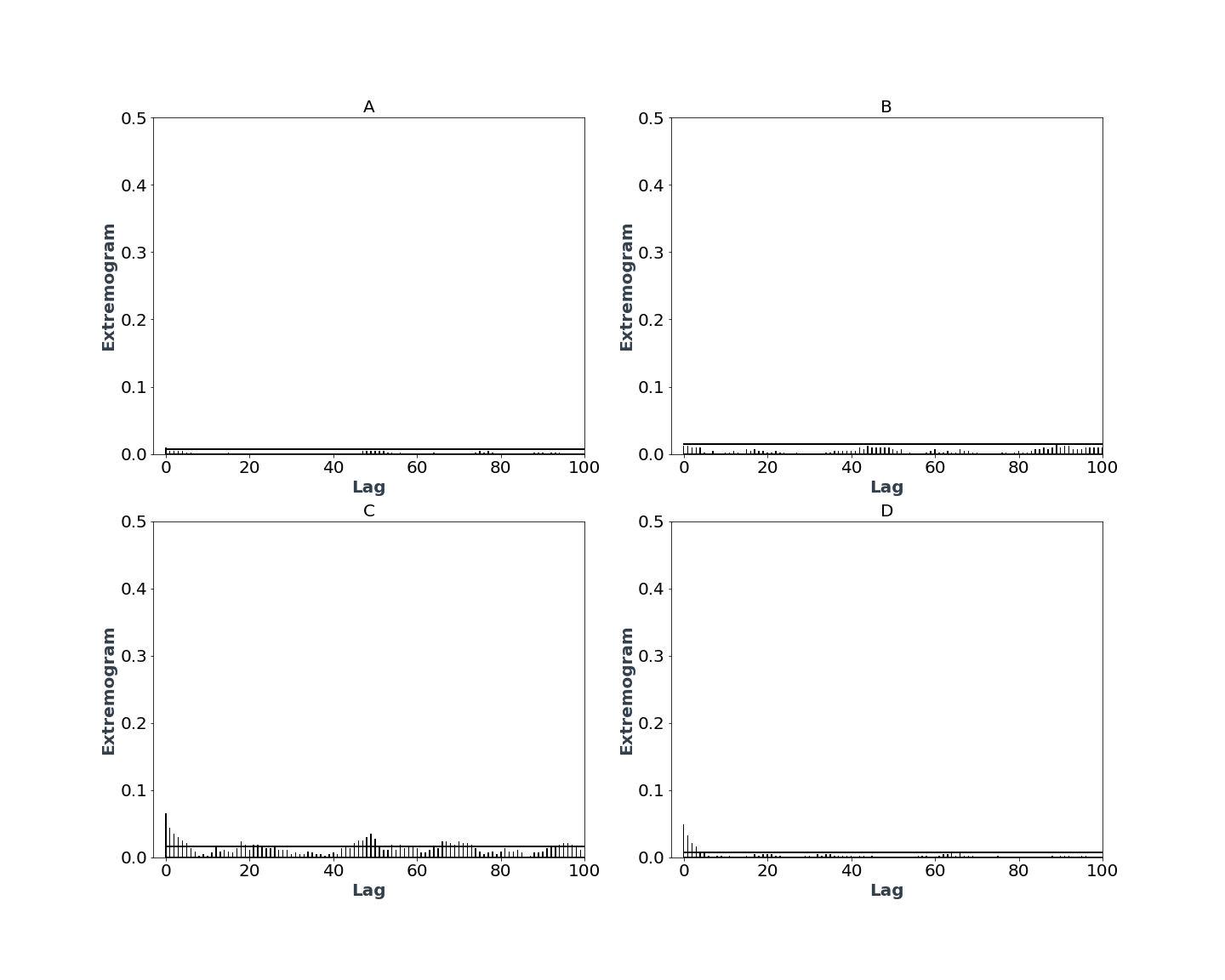}
			\caption[Cross-extremograms (half-hourly) conditioning on price spikes in TAS]{Sample cross-extremograms of half-hourly spot electricity prices conditioning on price spikes in TAS. The plot provides cross-extremograms for (A) TAS $\rightarrow$ NSW, (B) TAS $\rightarrow$ QLD, (C) TAS $\rightarrow$ SA and (D) TAS $\rightarrow$ VIC based on a sample period from 1 July 2009 to 30 June 2018.}
				\label{fig:cross_ext_TAS}
\end{figure}

In \Cref{fig:cross_ext_TAS}, significant cross-extremograms conditioning on price spikes in TAS are observed for SA and VIC. In both cases the extremal dependence decays to be insignificant within approximately five lags. This indicates the relatively low price influence and demand pressure from TAS on other markets.

\begin{figure}[!htbp]
			\centering
			\includegraphics[width=\linewidth,height=15cm]{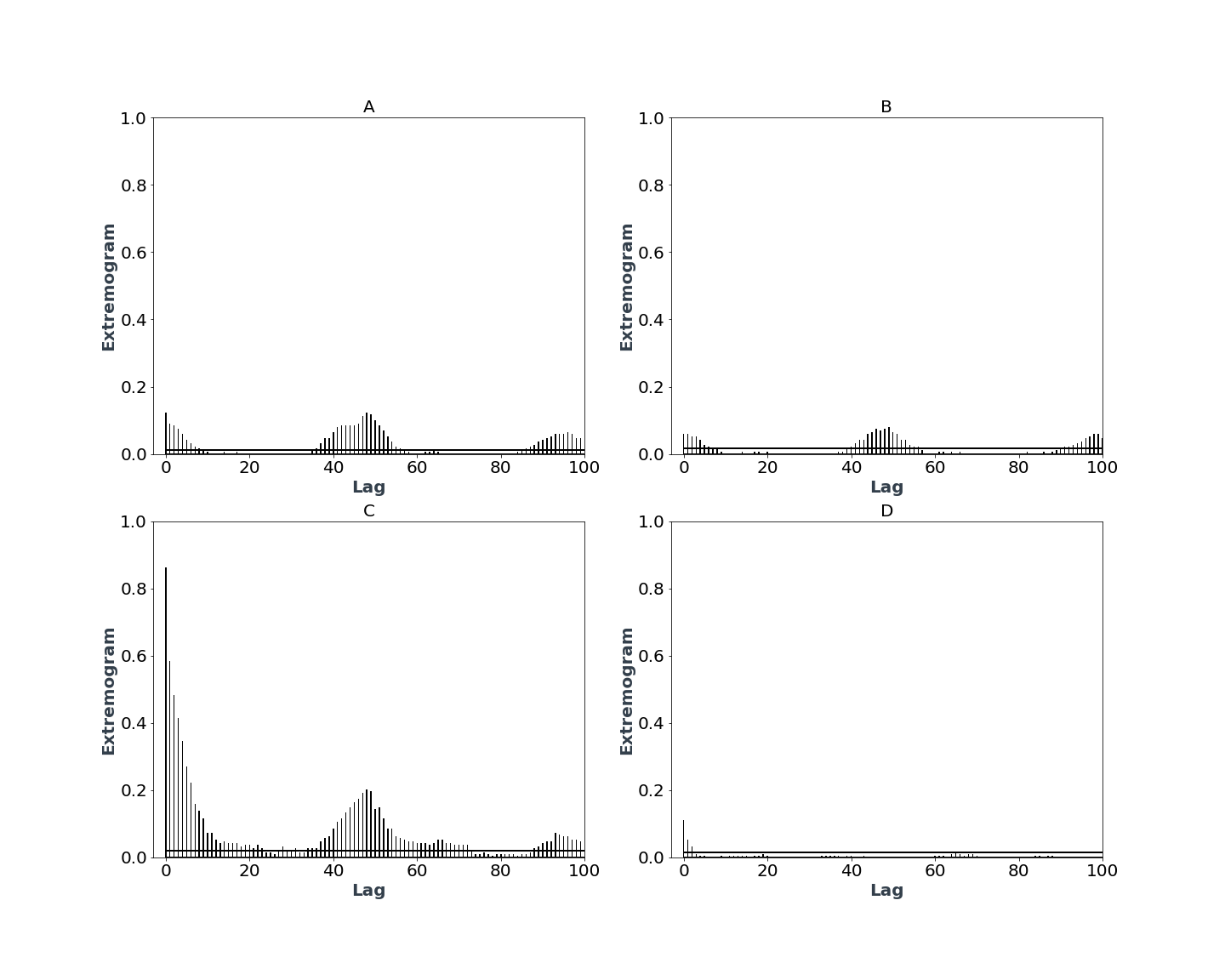}
			\caption[Cross-extremograms (half-hourly) conditioning on price spikes in VIC]{Sample cross-extremograms of half-hourly spot electricity prices conditioning on price spikes in VIC. The plot provides cross-extremograms for (A) VIC $\rightarrow$ NSW, (B) VIC $\rightarrow$ QLD, (C) VIC $\rightarrow$ SA and (D) VIC $\rightarrow$ TAS based on a sample period from 1 July 2009 to 30 June 2018. Note that the y-axis limits in this figure are different from those in \Cref{fig:cross_ext_NSW,fig:cross_ext_QLD,fig:cross_ext_SA,fig:cross_ext_TAS} due to high extremograms from VIC to SA.}
				\label{fig:cross_ext_VIC}
\end{figure}

\Cref{fig:cross_ext_VIC} illustrates that the VIC market has significant extremal dependence with all other considered markets; and the dependence between VIC and SA is the strongest. Note that in this figure, the cross-extremogram for SA at lag $0$, which is higher than 0.8, is significantly larger than the cross-extremogram for any other bi-variate relationship displayed in \Cref{fig:cross_ext_NSW,fig:cross_ext_QLD,fig:cross_ext_SA,fig:cross_ext_TAS,fig:cross_ext_VIC}. It indicates a large probability of joint extreme prices occurring contemporaneously in SA and VIC. This strong extremal dependence diminishes rapidly, while weaker dependence reappears around the $48^{th}$ lag.
On the other hand, cross-extremograms from VIC to NSW and QLD are more elevated around the $48^{th}$ lag than at the first several lags. In other words, given that there is a price spike in VIC at time $t$, e.g., today at a specific hour, the probability of observing a spike in NSW and QLD on the following day at $t+48$ is higher than the probability of contemporaneous spikes.
For the impact of extreme price events in VIC on NSW and SA, our findings may be explained the following way: since SA is only connected to VIC, it is more likely to be impacted directly by high prices in VIC. NSW, however, is also connected to QLD such that shocks to the VIC market might not transmit immediately to NSW. However, market expectations for the same hour on the following day then seem to significantly increase the likelihood of also observing a price spike in the NSW and QLD markets.
Interestingly, despite the one interconnector between VIC and TAS, the extremal dependence between these two markets appears to be weak and decays to be insignificant within five lags.

\newpage

\subsubsection{Analysis of 5-minute prices}
When 5-minute prices and an A\$5000/MWh threshold are considered (\Crefrange{fig:cross_ext_NSW_5min}{fig:cross_ext_VIC_5min}), the estimated extremograms show similar results to those observed for half-hourly data.\footnote{Note that we decided to move the cross-extremograms for 5-minute prices to the Appendix.} However, the magnitude of the estimated extremograms is typically lower, because there are less events where spot prices exceed the A\$5000/MWh threshold.
We find that there is significant extremal dependence between NSW and QLD and SA. Price spikes in NSW are likely to transmit to QLD both in subsequent periods and at the same time of the following day ($t+288$), while the extremograms from NSW to SA decay to be insignificant within 30 lags (2.5 hours).
QLD only has significant cross-extremograms with NSW, reflecting the direct interconnection between these two markets.
Cross-extremograms from SA to NSW and QLD are significant around the same time of the following day ($t+288$), based on a price spike in SA at period $t$. While there are no interconnectors between SA and either NSW or QLD, the detected extremal dependence reflects that SA is the most spiky and volatile market and tends to transmit some uncertainties to the entire NEM.
TAS exhibits no significant impact of extreme price shocks on other markets. This is partially due to the less extreme price observations in TAS, and also indicates a relatively low influence of TAS on the remaining NEM regions.
Regarding VIC, spikes in the cross-extremograms from VIC to SA are still the highest compared to those of other markets. Compared to the level of extremal dependence for the half-hourly analysis, 5-minute extremograms from VIC to all other markets are weaker, because VIC has the lowest mean spot price and the frequency of price spikes higher than A\$5000/MWh in VIC is lower than in NSW, QLD and SA.

Overall, applying cross-extremograms to both half-hourly and 5-minute data, we find the strongest extremal dependence between adjoining regional markets that are well connected through interconnectors, such as the pairs of NSW and QLD with two interconnectors, NSW and VIC with one interconnector, and SA and VIC with two interconnectors. This indicates that these pairs of markets have a strong tendency to exhibit extreme prices contemporaneously or jointly with certain time lags, i.e. strong and persistent spillover effects of price spikes.
On the other hand, significantly lower extremal dependence is observed between prices of the markets that are geographically distant and not directly interconnected, such as QLD and SA, QLD and VIC, and SA and TAS.
Interestingly, despite the existing \emph{Basslink} interconnector between VIC and TAS, the extremal dependence between prices in these two regional markets is relatively low.
This could be due to the different operating strategy of \emph{Basslink} as a merchant interconnector, compared to the other interconnectors which are regulated and government owned.
Meanwhile, the substantially different generation mix in TAS is also a contributing factor which leads to less price transmission effects between TAS and other regional markets.
Electricity in TAS is predominately generated from hydro power, while in the other markets, electricity generation is dominated by fossil fuel fired power plants such as coal and gas, or by intermittent generation from wind in SA.

In addition, the cross-extremograms between markets appear to be asymmetric. For example, conditioning on an extreme price in NSW, the probability of price spikes happening to VIC is generally insignificant except at short lags 0, 1, 2 and 3 (\Cref{fig:cross_ext_NSW}, Panel D). In contrast, conditioning on an extreme price in VIC, significant probabilities of price spikes happening to NSW can be observed both contemporaneously and around the $48^{th}$ and $96^{th}$ lags, illustrating more persistent extremal dependence (\Cref{fig:cross_ext_VIC}, Panel A). Given the highest level of interconnection to the rest of the NEM, the high local generation capacity from brown coal with relatively low costs, and the large export ratio of VIC, this asymmetry in cross-extremograms reflects the higher ability of VIC to transmit out local extreme prices compared to that of NSW.
Among all regions, the extreme price transmission from VIC to SA is the strongest. This reflects the fact that SA is highly reliant on importing electricity from VIC to firm its local supply because of the large share of wind generation in SA and the high uncertainty of wind output due to its intermittent nature.

\section{Impact of the 2016 Rebidding Rule Change}\label{section: rebidding casestudy}
This section applies extremograms to examine the impact of a particular market policy change -- the Australian Energy Market Commission (AEMC)'s rebidding rule change \citep{australian2015final} implemented on 1 July 2016.

The existence of strategic bidding and rebidding behaviour in the Australian NEM has been documented in the literature and various industry reports, see, for example, \cite{clements2016strategic}, \cite{dungey2018strategic} and \cite{wood2018mostly}.
The strategic behaviour is closely related to the current market setting, in particular, the half-hourly settlement and rebidding mechanism.
As introduced in \Cref{facts}, the settlement prices (i.e. trading prices) that generators are paid are the half-hourly average prices, no matter in which five-minute period the generators dispatch their output. This means that as long as there has been a 5-minute price spike, it is likely to increase the profit for the entire dispatch of a generator within the entire half-hour interval.
In addition, since generators in the NEM are allowed to rebid at any time up to five minutes before dispatch, there is an opportunity for generators to obtain priority to dispatch and thus secure their profit from their manufactured extreme prices.
For example, \cite{biggar2011theory} and \cite{clements2016strategic} provide empirical evidence on the attempt of baseload generators, who typically have higher market power and influence on the overall market supply and price, to maximise profit by combining strategic bidding and rebidding.
In particular, these generators first push up the dispatch price by withholding capacity at low price levels and meanwhile bidding the remaining capacity at the price cap.
After creating a price spike in the first 5-minute dispatch interval, these generators then add capacity at the lowest price levels through the rebidding mechanism to ensure their dispatch priority within the same half-hourly interval, before their competitors are able to respond to the price change.

Strategic bidding and rebidding behaviour can result in various undesirable consequences for the market. As argued in \cite{clements2016strategic}, \cite{hurn2016smooth} and \cite{dungey2018strategic}, strategic behaviour in electricity markets leads to extreme price events which do not reflect the fundamental generation cost, inflate final electricity settlement prices and significantly increase market volatility. Furthermore, \cite{clements2016strategic} provide empirical evidence that strategic bidding substantially changes regional market supply conditions and can cause market anomalies such as electricity flowing from a high price area to a low price area (i.e. the so-called counter-price flow).
Due to the concerns about the negative impacts of strategic rebidding, the AEMC implemented a rule change in 2016 which replaced the requirement that `offers be made in good faith' by a prohibition against making false or misleading offers and put stronger restrictions on `last minute' electricity market rebids. In particular, a new definition of the late rebidding period was inserted: the period beginning fifteen minutes before the commencement of a half-hourly trading interval and ending at the end of this trading interval is a late rebidding period.
The new rule also added a requirement on information disclosure of late rebids, that is, all rebids that are close to dispatch time have to be submitted with a reason and contemporaneous record of the circumstances that led to the necessity of late rebids. Furthermore, the rule change added restrictions on the timing of rebidding, i.e., any variation has to be made as soon as practicable after a change in material circumstances and conditions without delay.

Strategic behaviour of generators and the rebidding rule change are relevant to our examination of extremal price dependence because,
according to \cite{hurn2016smooth}, isolated price spikes are more likely to be caused by strategic behaviour than those spikes continuing for more consecutive periods.
In particular, artificial spikes can be eliminated within a short period as soon as other market competitors respond to it by adding supply, while extreme events led by real supply shortage (e.g. network outage or high demand) will persist until the underlying supply issue is resolved.
Since the rule change with more restrictive timing and disclosure requirement is expected to reduce strategic behaviour, price spikes in the post-rule-change period are expected to be more persistent than those in the pre-rule-change period. We therefore apply extremograms to validate this hypothesis and expect to see more isolated spikes in the pre-implementation period as well as more persistence in the post-implementation period.

\begin{figure}[!h]
	\centering
	\begin{minipage}[t]{1\linewidth}
		\includegraphics[width = 1\linewidth, height=6cm]{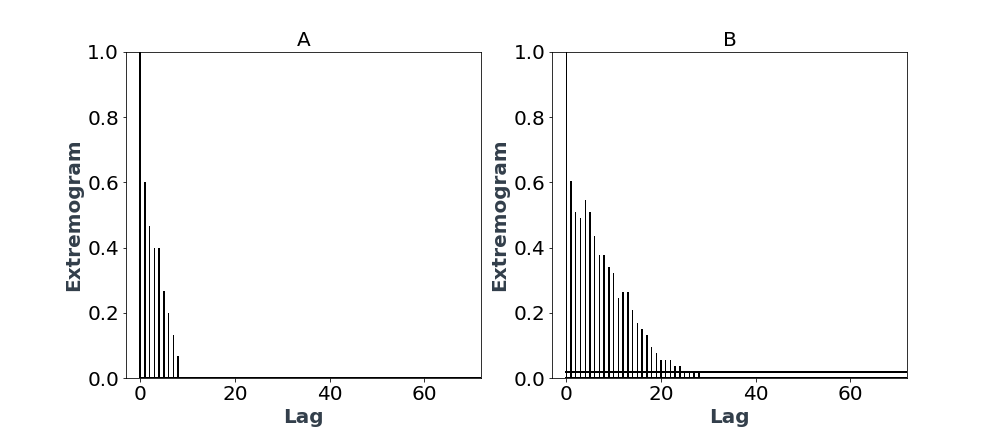}
		\subcaption*{\footnotesize{A: NSW}}
		\vspace{1em}
	\end{minipage}
	\begin{minipage}[t]{1\linewidth}	
		\includegraphics[width = 1\linewidth, height=6cm]{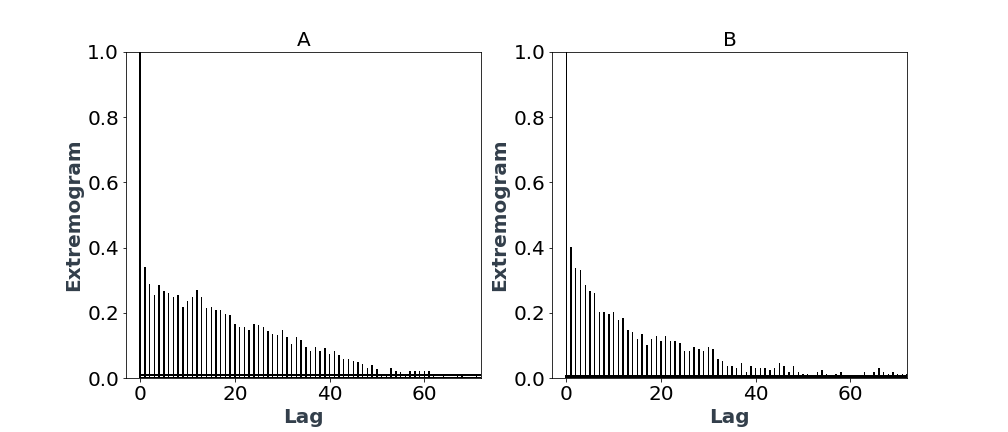}
		\subcaption*{\footnotesize{B: QLD}}
	\end{minipage}
	\begin{minipage}[t]{1\linewidth}	
		\includegraphics[width = 1\linewidth, height=6cm]{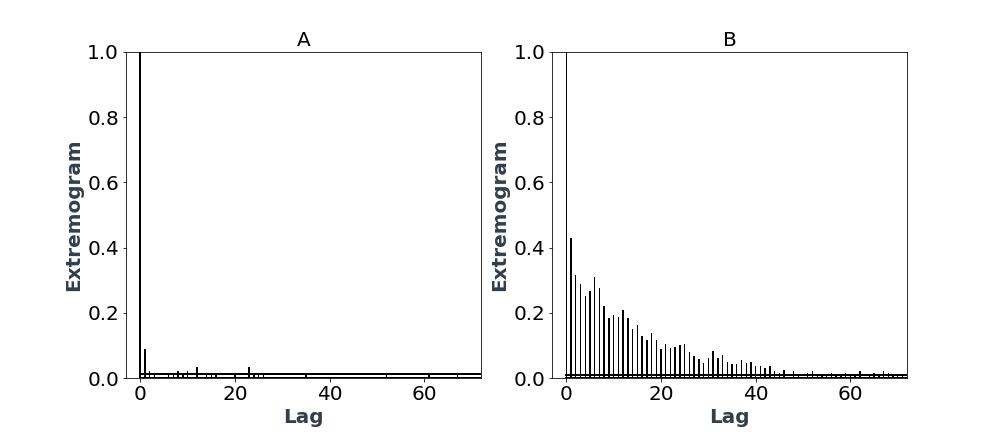}
		\subcaption*{\footnotesize{C: SA}}
	\end{minipage}	
	\caption[Univariate extremograms pre- and post- the 2016 rule change (NSW, QLD and SA)]{Univariate extremograms pre- and post- the rebidding rule change on 1 July 2016 for (A) NSW, (B) QLD and (C) SA. Pre-rule-change period is from 1 July 2014 to 30 June 2016. Post-rule-change period is from 1 July 2016 to 30 June 2018.}
	\label{fig:rule_change1}
\end{figure}

\begin{figure}[!h]
	\centering
	\begin{minipage}[t]{1\linewidth}
		\includegraphics[width = 1\linewidth, height=6cm]{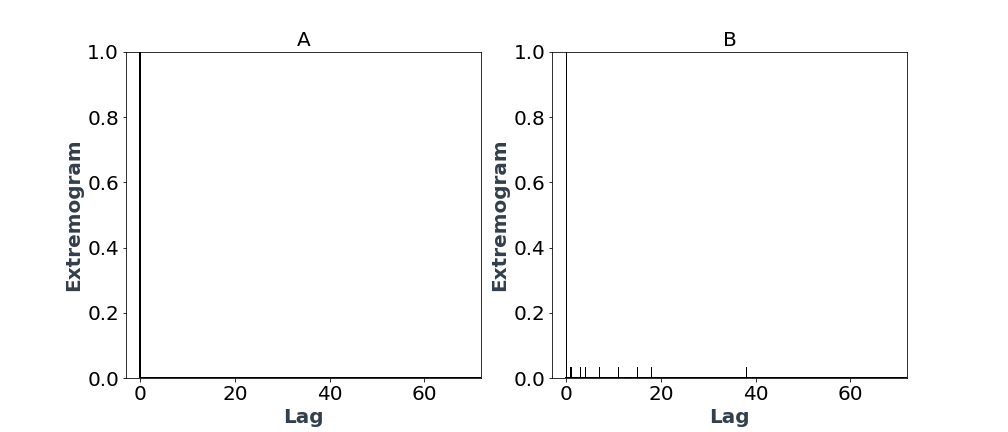}
		\subcaption*{\footnotesize{A: TAS}}
		\vspace{1em}
	\end{minipage}
	\begin{minipage}[t]{1\linewidth}	
		\includegraphics[width = 1\linewidth, height=6cm]{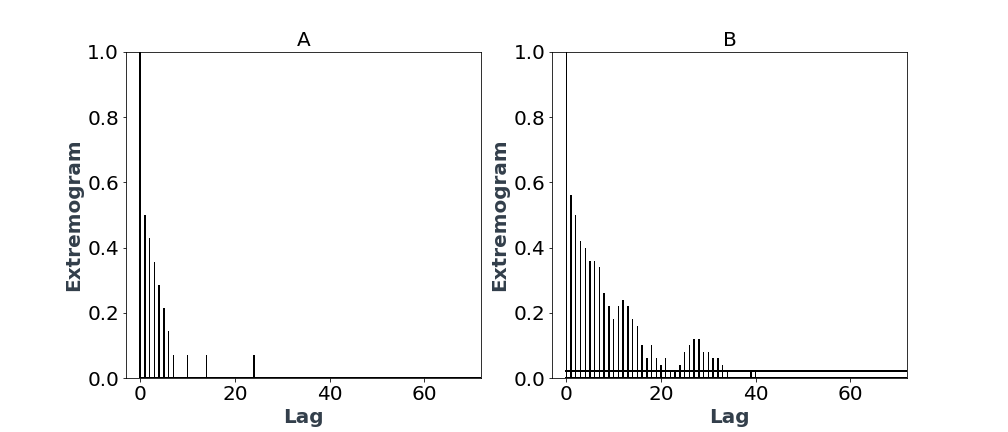}
		\subcaption*{\footnotesize{B: VIC}}
	\end{minipage}
	
	\caption[Univariate extremograms pre- and post- the 2016 rule change (TAS and VIC)]{Univariate extremograms pre- and post- the rebidding rule change on 1 July 2016 for (A) TAS and (B) VIC. Pre-rule-change period is from 1 July 2014 to 30 June 2016. Post-rule-change period is from 1 July 2016 to 30 June 2018.}
	\label{fig:rule_change2}
\end{figure}

To examine the impact of the AEMC's rule change on the persistence of price spikes in the NEM, we estimate univariate extremograms for the 2-year pre-rule-change period (1 July 2014 to 30 June 2016) and the 2-year post-rule-change period (1 July 2016 to 30 June 2018).
The estimated extremograms are based on 5-minute dispatch prices, because, compared to 30-minute prices which have been smoothed and possibly lowered by averaging across six 5-minute price intervals, 5-minute prices directly capture all price spikes and are able to reflect the persistence of extreme price events.
Meanwhile, the A\$5000/MWh threshold is used because, in order for generators to obtain a high half-hourly average price in the above described strategy, the price spike they create in a single 5-minute interval needs to be high enough to remain elevated after possibly being averaged with five low prices from other dispatch intervals\footnote{
For illustration, assume that a price spike at A\$6000/MWh occurs in the first 5-minute dispatch interval. Then even if the price drops to zero (i.e. (A\$0)/MWh) for the remaining time of the whole 30-minute trading interval, the trading price is A\$$(\frac{6000+ 5 * (0)}{6} = 1000)$, which is significantly higher than the typical spot prices in the NEM under normal conditions.}.
The results are presented in \Cref{fig:rule_change1,fig:rule_change2}. The figures illustrate that extreme prices appear to be more persistent in the post-rule-change period than in the pre-rule-change period to the extent that significant extremograms can be found in more consecutive lags in the post period. The increase in persistence is especially apparent for SA in Panel C of \Cref{fig:rule_change1}. Specifically, while there is only 10\% probability of another spike happening in the next 5-minute interval given a price spike in SA in the pre-implementation period, in the post-implementation period the probability of a spike directly following another is as high as 42\%. Furthermore, while extremograms for SA in the pre-implementation period decay to be insignificant rapidly after one lag (i.e. 5 minutes), extremograms in the post-implementation period become insignificant far more slowly, taking around 50 lags (approximately 4 hours).

In addition to the extremogram analysis in \Cref{fig:rule_change1,fig:rule_change2}, \Cref{Table: Pre and post spikes} describes the occurrence distribution of price spikes in order to validate the increased spike persistence indicated by the extremogram analysis. In particular, \Cref{Table: Pre and post spikes} counts the total occurrence of prices greater than A\$5000/MWh for both the pre- and post-implementation period and reports the number of half-hour intervals in which these spikes occur. The `Spikes per Interval' column calculates the average spike count in a `spiky' half-hourly interval.
According to the table, firstly, except for QLD, there are more spikes observed in the post-implementation period. This is partly due to the higher volatility of the market itself in recent years, because of the exit of some major fossil-fuel fired plants and the increasing share of intermittent renewable generations \citep{rai2019australia}. Recall that our focus is on the clustering or persistence of spikes (i.e. how close or how distant the spikes are to each other) rather than the total number of spikes. Although the number of half-hourly spiky intervals has also increased for each market except for QLD, the extent of this increase is less than the increase in total spike counts. In NSW, for example, where the total spike count has increased three times in the post-implementation period compared to the pre-implementation period, the number of spiky half-hourly intervals has increased only about two times. As a result, the spike count per half-hourly interval (i.e. spikes happening in the same half-hourly interval) in the post-implementation period is higher than that in the pre-implementation period for all markets. This suggests that on average more spike clustering and persistence is observed in the NEM, while we observe less `isolated' spikes that could be a sign of strategic bidding or rebidding.

In addition, the `Events Count' column in \Cref{Table: Pre and post spikes} shows the count of physical markets events relevant to price spikes during the pre-rule-change and post-rule-change period, respectively. Market events data are collected from AEMO's published market notices (\url{https://aemo.com.au/market-notices}) on extraordinary market issues including demand shocks, short-term generation outage, transmission constraints. The results show that spikes in the post period are more likely to be related to significant physical market events.
These results validate the above findings for the estimated extremograms: in the post period, a larger proportion of spikes are persistent events rather than short-term or isolated `artificial' events. The persistent spikes are likely caused by physical power system issues which require more time to recover, while the latter are more likely to be the result of strategic bidding or rebidding.

\begin{table}[!h]
	\renewcommand\arraystretch{1.1}
	\centering
	\footnotesize
	\caption[Price spikes in the pre-rule-change period and post-rule-change period]{Price spikes which occurred in pre-rule-change period (1 July 2014 to 30 June 2016) and post-rule-change period (1 July 2016 to 30 June 2018). `Spike Count' refers to the total occurrence of prices greater than A\$5000/MWh; `Half-Hour Intervals' counts the number of half-hour intervals containing spikes; `Spikes per Interval' refers to the mean spike count in a `spiky' interval. For example, the first row in Panel (a) shows that there were 15 price spikes in NSW during the pre-rule-change period which occurred in 7 distinct half-hour periods. On average, each of these half-hour periods contains 2.14 spikes. `Events Count' shows the count of market events related to price spikes in the corresponding periods. Market events data are collected from AEMO's published market notices (\url{https://aemo.com.au/market-notices}) on extraordinary market issues including demand shocks, short-term generation outage, transmission constraints.}
	\label{Table: Pre and post spikes}
	\begin{tabular*}{1\textwidth}{@{\extracolsep{\fill}}lrrrr}
		\hline
		{Market}&Spike Count ($sc$)& Half-Hour Intervals ($hh$)& Spikes per Interval ($\frac{sc}{hh}$)&Events Count\\
		\hline
		
		\multicolumn{5}{l}{\emph{Panel (a) : Pre-rule-change period (1 July 2014 to 30 June 2016)}}\\
		NSW & 15 & 7 & 2.14 & 28\\
		QLD & 229 & 153 & 1.50 & 154\\
		SA & 88 & 80 & 1.10 & 64\\
		TAS & 10 & 10 & 1.00 & 6\\
		VIC & 14 & 8 & 1.75 & 46\\
		\multicolumn{5}{l}{}\\
		\multicolumn{5}{l}{\emph{Panel (b) : Post-rule-change period (1 July 2016 to 30 June 2018)}}\\
		NSW & 53 & 23 & 2.30 & 82\\
		QLD & 157 & 97 & 1.62 & 170\\
		SA & 236 & 142 & 1.66 & 420\\
		TAS & 27 & 25 & 1.08 & 12\\
		VIC & 48 & 22 & 2.18 & 90\\
		\hline
		
	\end{tabular*}
	\begin{flushleft}
	\end{flushleft}
\end{table}

In summary, this section applies extremograms to investigate the impact of the AEMC's 2016 rule change on generator rebidding. Our findings indicate a higher probability of observing isolated abnormally high price events in the pre-implementation period and, in comparison, a higher probability of observing persistent price spikes in the post-implementation period. This can be considered as an indication for less strategic behaviour after the rule change.

\section{Conclusions}\label{conclusions}

In this study we have examined extremal dependence between spot electricity prices in different regions in the Australian NEM. The risk of extreme outcomes and joint price spikes in regional markets is of significant concern to market participants and regulators.
So far only a limited number of studies have concentrated on modelling the dependence within a single or between different regional electricity markets with a particular focus on extreme observations. We apply a relatively new econometric technique, the extremogram developed by \cite{davis2009} and \cite{davis2011,davis2012} to generate important insights on the nature of extreme price outcomes and their transmission in interconnected spot electricity markets.

In particular, we examine the persistence of extreme price events, transmission effects of price spikes between different regions, and the impact of interconnection within the Australian NEM.
Applying the extremogram to both half-hourly and 5-minute prices from July 2009 to June 2018, we find that the extreme prices are most persistent in the market with the highest share of intermittent renewable energy, namely SA, but also in more concentrated markets such as QLD and TAS.
We argue that limited interconnection capacity, high price volatility caused by intermittent generation, high supply costs due to market power of several local generators, and the potential inability to import electricity from other regions to replace local supply, are contributing factors to the occurrence of extreme price outcomes. These factors also contribute to higher levels of persistence for extreme prices within a regional market.
The estimated extremal dependence for each region is typically stronger with a one-day time lag, which indicates that given an extreme price observation in an individual region, there is a reasonably high likelihood of another extreme outcome at around the same time on the next day.
We also find that extreme prices are transmitted between regions. These transmission effects are typically more pronounced between adjoining and physically interconnected markets and show asymmetric patterns depending on the relative importance and size of the regional markets.
Furthermore, the extent of extremal dependence in half-hourly prices above A\$300/WMh tends to be stronger than extremal dependence for 5-minute interval prices above A\$5000/MWh.

We further apply the extremogram to investigate the impact of a particular market event: the AEMC's 2016 rebidding rule change which was aimed at preventing strategic and misleading bidding behaviour.
Our results show higher persistence of extreme prices after the implementation of the rule change, which aligns better with spikes caused by `real' market events that impact on the supply-demand relationship. This indicates some effectiveness of the rule change in terms of reducing `isolated' price spikes that are more likely the outcome of strategic bidding behaviour of market participants.

Our study helps to provide a better understanding of the persistence of extreme price outcomes in individual electricity markets as well as the relationship between extreme events across regional electricity spot markets. This provides important insights for market participants, especially for those who simultaneously operate in different regional markets in the NEM, in making their planning, trading and risk management decisions. Our results are also important for regulators in reducing unnecessarily high prices and market volatility as well as improving the efficiency of electricity pricing through effective policy making and regulation of participant behaviours.

Finally, there are some directions for future research.
In the context of rapidly developing storage facilities and demand response mechanisms, the application of extremograms in demand management will be of interest to market participants. By incorporating the estimated extremal dependence of the spot price series, important implications can be gained on the optimal timing of operations including demand curtailment, the charging or discharging of large batteries, as well as for relevant revenue forecasting and the pricing of demand curtailment contracts.
Meanwhile, given various available financial derivatives related to electricity markets such as the `\$300 Cap Products' and interregional settlement residue auctions\footnote{For details of the interregional settlement residue auctions in the NEM, see https://aemo.com.au/en/energy-systems/electricity/national-electricity-market-nem/market-operations/settlements-and-payments/settlements/settlements-residue-auction.} which allow the future interregional price difference to be traded, application of extremograms in combination with derivatives to hedge the extreme price risk will also be valuable. In particular, relating the estimated extremograms of prices and the payoff distribution of a hedged portfolio will provide important insights for market participants for effective risk management.
Furthermore, the sample extremogram provides important information for model selection by assessing the time series patterns of sample data. Evaluating the performance of an econometric framework in electricity markets, such as, e.g., GARCH type or regime-switching models, model selection informed by extremograms may also provide interesting evidence on the effectiveness of extremograms on capturing particular features in electricity price data.


\newpage
\setstretch{1}
\renewcommand\bibname{References}
\bibliographystyle{elsarticle-harv}
\bibliography{extremogram_bib}

\begin{thebibliography}{72}
\expandafter\ifx\csname natexlab\endcsname\relax\def\natexlab#1{#1}\fi
\providecommand{\url}[1]{\texttt{#1}}
\providecommand{\href}[2]{#2}
\providecommand{\path}[1]{#1}
\providecommand{\DOIprefix}{doi:}
\providecommand{\ArXivprefix}{arXiv:}
\providecommand{\URLprefix}{URL: }
\providecommand{\Pubmedprefix}{pmid:}
\providecommand{\doi}[1]{\href{http://dx.doi.org/#1}{\path{#1}}}
\providecommand{\Pubmed}[1]{\href{pmid:#1}{\path{#1}}}
\providecommand{\bibinfo}[2]{#2}
\ifx\xfnm\relax \def\xfnm[#1]{\unskip,\space#1}\fi
\bibitem[{Aderounmu and Wolff(2014)}]{aderounmu2014modeling}
\bibinfo{author}{Aderounmu, A.A.}, \bibinfo{author}{Wolff, R.},
  \bibinfo{year}{2014}.
\newblock \bibinfo{title}{{Modeling dependence of price spikes in Australian
  electricity markets}}.
\newblock \bibinfo{journal}{Energy Risk} \bibinfo{volume}{11},
  \bibinfo{pages}{60--65}.
\bibitem[{Apergis et~al.(2016)Apergis, Fontini and
  Inchauspe}]{apergis2016integration}
\bibinfo{author}{Apergis, N.}, \bibinfo{author}{Fontini, F.},
  \bibinfo{author}{Inchauspe, J.}, \bibinfo{year}{2016}.
\newblock \bibinfo{title}{{Integration of regional electricity markets in
  Australia: A price convergence assessment}}.
\newblock \bibinfo{journal}{Energy Economics} \bibinfo{volume}{52},
  \bibinfo{pages}{176--182}.
\bibitem[{{ASX Limited}(2019)}]{ASX2019contract}
\bibinfo{author}{{ASX Limited}}, \bibinfo{year}{2019}.
\newblock \bibinfo{title}{Contract specifications}.
\newblock \URLprefix
  \url{{https://www.asx.com.au/documents/products/asx24-contract-specifications.pdf}}.
  \bibinfo{note}{accessed 30 May 2020}.
\bibitem[{{Australian Energy Market Commission}(2015)}]{australian2015final}
\bibinfo{author}{{Australian Energy Market Commission}}, \bibinfo{year}{2015}.
\newblock \bibinfo{title}{Final rule determination national electricity
  amendment (bidding in good faith)}.
\newblock \URLprefix
  \url{{https://www.aemc.gov.au/rule-changes/bidding-in-good-faith}}.
  \bibinfo{note}{accessed 1 May 2020}.
\bibitem[{{Australian Energy Regulator}(2018)}]{AEMO2014}
\bibinfo{author}{{Australian Energy Regulator}}, \bibinfo{year}{2018}.
\newblock \bibinfo{title}{State of the energy market}.
\newblock \URLprefix
  \url{https://www.aer.gov.au/publications/state-of-the-energy-market-reports/state-of-the-energy-market-2018}.
  \bibinfo{note}{accessed 21 October 2018}.
\bibitem[{{Australian Energy Regulator}(2020)}]{aer15}
\bibinfo{author}{{Australian Energy Regulator}}, \bibinfo{year}{2020}.
\newblock \bibinfo{title}{Wholesale statistics: {Quarterly} interregional trade
  as a percentage of regional energy consumption}.
\newblock \URLprefix
  \url{https://www.aer.gov.au/wholesale-markets/wholesale-statistics/quarterly-interregional-trade-as-a-percentage-of-regional-energy-consumption}.
  \bibinfo{note}{accessed 30 May 2020}.
\bibitem[{Bierbrauer et~al.(2007)Bierbrauer, Menn, Rachev and
  Tr\"{u}ck}]{BieMennRachevTrueck}
\bibinfo{author}{Bierbrauer, M.}, \bibinfo{author}{Menn, C.},
  \bibinfo{author}{Rachev, S.}, \bibinfo{author}{Tr\"{u}ck, S.},
  \bibinfo{year}{2007}.
\newblock \bibinfo{title}{Spot and derivative pricing in the {EEX} power
  market}.
\newblock \bibinfo{journal}{Journal of Banking \& Finance}
  \bibinfo{volume}{31}, \bibinfo{pages}{3462--3485}.
\bibitem[{Bigerna et~al.(2017)Bigerna, Bollino, Ciferri and
  Polinori}]{bigerna2017renewables}
\bibinfo{author}{Bigerna, S.}, \bibinfo{author}{Bollino, C.A.},
  \bibinfo{author}{Ciferri, D.}, \bibinfo{author}{Polinori, P.},
  \bibinfo{year}{2017}.
\newblock \bibinfo{title}{Renewables diffusion and contagion effect in
  {Italian} regional electricity markets: {Assessment} and policy
  implications}.
\newblock \bibinfo{journal}{Renewable and Sustainable Energy Reviews}
  \bibinfo{volume}{68}, \bibinfo{pages}{199--211}.
\bibitem[{Biggar(2011)}]{biggar2011theory}
\bibinfo{author}{Biggar, D.}, \bibinfo{year}{2011}.
\newblock \bibinfo{title}{The theory and practice of the exercise of market
  power in the {Australian NEM}}.
\newblock \bibinfo{journal}{Australian Competition and Consumer Commission,
  Melbourne}
  \DOIprefix\doi{http://citeseerx.ist.psu.edu/viewdoc/summary?doi=10.1.1.465.2444}.
\bibitem[{Bollerslev et~al.(2013)Bollerslev, Todorov and
  Li}]{bollerslev2013jump}
\bibinfo{author}{Bollerslev, T.}, \bibinfo{author}{Todorov, V.},
  \bibinfo{author}{Li, S.Z.}, \bibinfo{year}{2013}.
\newblock \bibinfo{title}{Jump tails, extreme dependencies, and the
  distribution of stock returns}.
\newblock \bibinfo{journal}{Journal of Econometrics} \bibinfo{volume}{172},
  \bibinfo{pages}{307--324}.
\bibitem[{Cho et~al.(2016)Cho, Davis and Ghosh}]{cho2016asymptotic}
\bibinfo{author}{Cho, Y.B.}, \bibinfo{author}{Davis, R.A.},
  \bibinfo{author}{Ghosh, S.}, \bibinfo{year}{2016}.
\newblock \bibinfo{title}{Asymptotic properties of the empirical spatial
  extremogram}.
\newblock \bibinfo{journal}{Scandinavian Journal of Statistics}
  \bibinfo{volume}{43}, \bibinfo{pages}{757--773}.
\bibitem[{Christensen et~al.(2009)Christensen, Hurn and
  Lindsay}]{christensen2009never}
\bibinfo{author}{Christensen, T.}, \bibinfo{author}{Hurn, S.},
  \bibinfo{author}{Lindsay, K.}, \bibinfo{year}{2009}.
\newblock \bibinfo{title}{It never rains but it pours: modeling the persistence
  of spikes in electricity prices}.
\newblock \bibinfo{journal}{The Energy Journal} \bibinfo{volume}{30},
  \bibinfo{pages}{25--48}.
\bibitem[{Christensen et~al.(2012)Christensen, Hurn and
  Lindsay}]{christensen2012forecasting}
\bibinfo{author}{Christensen, T.M.}, \bibinfo{author}{Hurn, A.S.},
  \bibinfo{author}{Lindsay, K.A.}, \bibinfo{year}{2012}.
\newblock \bibinfo{title}{Forecasting spikes in electricity prices}.
\newblock \bibinfo{journal}{International Journal of Forecasting}
  \bibinfo{volume}{28}, \bibinfo{pages}{400--411}.
\bibitem[{Ciarreta and Zarraga(2015)}]{ciarreta2015analysis}
\bibinfo{author}{Ciarreta, A.}, \bibinfo{author}{Zarraga, A.},
  \bibinfo{year}{2015}.
\newblock \bibinfo{title}{{Analysis of mean and volatility price transmissions
  in the MIBEL and EPEX electricity spot markets}}.
\newblock \bibinfo{journal}{Energy Journal} \bibinfo{volume}{36},
  \bibinfo{pages}{41--60}.
\bibitem[{Clements et~al.(2013)Clements, Fuller and Hurn}]{clements2013semi}
\bibinfo{author}{Clements, A.}, \bibinfo{author}{Fuller, J.},
  \bibinfo{author}{Hurn, S.}, \bibinfo{year}{2013}.
\newblock \bibinfo{title}{Semi-parametric forecasting of spikes in electricity
  prices}.
\newblock \bibinfo{journal}{Economic Record} \bibinfo{volume}{89},
  \bibinfo{pages}{508--521}.
\bibitem[{Clements et~al.(2015)Clements, Herrera and
  Hurn}]{clements2015modelling}
\bibinfo{author}{Clements, A.}, \bibinfo{author}{Herrera, R.},
  \bibinfo{author}{Hurn, A.}, \bibinfo{year}{2015}.
\newblock \bibinfo{title}{Modelling interregional links in electricity price
  spikes}.
\newblock \bibinfo{journal}{Energy Economics} \bibinfo{volume}{51},
  \bibinfo{pages}{383--393}.
\bibitem[{Clements et~al.(2016)Clements, Hurn and Li}]{clements2016strategic}
\bibinfo{author}{Clements, A.}, \bibinfo{author}{Hurn, A.},
  \bibinfo{author}{Li, Z.}, \bibinfo{year}{2016}.
\newblock \bibinfo{title}{Strategic bidding and rebidding in electricity
  markets}.
\newblock \bibinfo{journal}{Energy Economics} \bibinfo{volume}{59},
  \bibinfo{pages}{24--36}.
\bibitem[{Cont and Kan(2011)}]{cont2011statistical}
\bibinfo{author}{Cont, R.}, \bibinfo{author}{Kan, Y.H.G.},
  \bibinfo{year}{2011}.
\newblock \bibinfo{title}{Statistical modeling of credit default swap
  portfolios}.
\newblock \bibinfo{journal}{Available at SSRN}
  \DOIprefix\doi{https://dx.doi.org/10.2139/ssrn.1771862}.
\bibitem[{Davis and Mikosch(2009)}]{davis2009}
\bibinfo{author}{Davis, R.A.}, \bibinfo{author}{Mikosch, T.},
  \bibinfo{year}{2009}.
\newblock \bibinfo{title}{The extremogram: {A} correlogram for extreme events}.
\newblock \bibinfo{journal}{Bernoulli} \bibinfo{volume}{15},
  \bibinfo{pages}{977--1009}.
\bibitem[{Davis et~al.(2011)Davis, Mikosch and Cribben}]{davis2011}
\bibinfo{author}{Davis, R.A.}, \bibinfo{author}{Mikosch, T.},
  \bibinfo{author}{Cribben, I.}, \bibinfo{year}{2011}.
\newblock \bibinfo{title}{Estimating extremal dependence in univariate and
  multivariate time series via the extremogram}.
\newblock \bibinfo{journal}{arXiv:1107.5592}
  \DOIprefix\doi{https://arxiv.org/abs/1107.5592}.
\bibitem[{Davis et~al.(2012)Davis, Mikosch and Cribben}]{davis2012}
\bibinfo{author}{Davis, R.A.}, \bibinfo{author}{Mikosch, T.},
  \bibinfo{author}{Cribben, I.}, \bibinfo{year}{2012}.
\newblock \bibinfo{title}{Towards estimating extremal serial dependence via the
  bootstrapped extremogram}.
\newblock \bibinfo{journal}{Journal of Econometrics} \bibinfo{volume}{170},
  \bibinfo{pages}{142--153}.
\bibitem[{Davis et~al.(2013)Davis, Mikosch and Zhao}]{davis2013measures}
\bibinfo{author}{Davis, R.A.}, \bibinfo{author}{Mikosch, T.},
  \bibinfo{author}{Zhao, Y.}, \bibinfo{year}{2013}.
\newblock \bibinfo{title}{Measures of serial extremal dependence and their
  estimation}.
\newblock \bibinfo{journal}{Stochastic Processes and their Applications}
  \bibinfo{volume}{123}, \bibinfo{pages}{2575--2602}.
\bibitem[{De~Menezes and Houllier(2015)}]{de2015germany}
\bibinfo{author}{De~Menezes, L.M.}, \bibinfo{author}{Houllier, M.A.},
  \bibinfo{year}{2015}.
\newblock \bibinfo{title}{{Germany's nuclear power plant closures and the
  integration of electricity markets in Europe}}.
\newblock \bibinfo{journal}{Energy Policy} \bibinfo{volume}{85},
  \bibinfo{pages}{357--368}.
\bibitem[{De~Menezes and Houllier(2016)}]{de2014reassessing}
\bibinfo{author}{De~Menezes, L.M.}, \bibinfo{author}{Houllier, M.A.},
  \bibinfo{year}{2016}.
\newblock \bibinfo{title}{{Reassessing the integration of European electricity
  markets: A fractional cointegration analysis}}.
\newblock \bibinfo{journal}{Energy Economics} \bibinfo{volume}{53},
  \bibinfo{pages}{132--150}.
\bibitem[{De~Vany and Walls(1999a)}]{devany99}
\bibinfo{author}{De~Vany, A.}, \bibinfo{author}{Walls, W.},
  \bibinfo{year}{1999}a.
\newblock \bibinfo{title}{Cointegration analysis of spot electricity prices:
  insights on transmission efficiency in the {W}estern {US}}.
\newblock \bibinfo{journal}{Energy Economics} \bibinfo{volume}{21(5)},
  \bibinfo{pages}{435--448}.
\bibitem[{De~Vany and Walls(1999b)}]{de1999price}
\bibinfo{author}{De~Vany, A.S.}, \bibinfo{author}{Walls, W.D.},
  \bibinfo{year}{1999}b.
\newblock \bibinfo{title}{Price dynamics in a network of decentralized power
  markets}.
\newblock \bibinfo{journal}{Journal of Regulatory Economics}
  \bibinfo{volume}{15}, \bibinfo{pages}{123--140}.
\bibitem[{Dempster et~al.(2008)Dempster, Isaacs and Smith}]{Dempster08}
\bibinfo{author}{Dempster, G.}, \bibinfo{author}{Isaacs, J.},
  \bibinfo{author}{Smith, N.}, \bibinfo{year}{2008}.
\newblock \bibinfo{title}{Price discovery in restructured electricity markets}.
\newblock \bibinfo{journal}{Resource and Energy Economics}
  \bibinfo{volume}{30(2)}, \bibinfo{pages}{250--259}.
\bibitem[{Dickey and Fuller(1979)}]{dickey1979distribution}
\bibinfo{author}{Dickey, D.A.}, \bibinfo{author}{Fuller, W.A.},
  \bibinfo{year}{1979}.
\newblock \bibinfo{title}{Distribution of the estimators for autoregressive
  time series with a unit root}.
\newblock \bibinfo{journal}{Journal of the American Statistical Association}
  \bibinfo{volume}{74}, \bibinfo{pages}{427--431}.
\bibitem[{Do et~al.(2020)Do, Nepal and Smyth}]{do2020interconnectedness}
\bibinfo{author}{Do, H.X.}, \bibinfo{author}{Nepal, R.},
  \bibinfo{author}{Smyth, R.}, \bibinfo{year}{2020}.
\newblock \bibinfo{title}{Interconnectedness in the {Australian National
  Electricity Market}: A higher moment analysis}.
\newblock \bibinfo{journal}{Economic Record} \bibinfo{volume}{315},
  \bibinfo{pages}{450--469}.
\bibitem[{Drees(2015)}]{drees2015bootstrapping}
\bibinfo{author}{Drees, H.}, \bibinfo{year}{2015}.
\newblock \bibinfo{title}{Bootstrapping empirical processes of cluster
  functionals with application to extremograms}.
\newblock \bibinfo{journal}{arXiv:1511.00420}
  \DOIprefix\doi{https://arxiv.org/abs/1511.00420}.
\bibitem[{Dungey et~al.(2018)Dungey, Ghahremanlou and
  Long}]{dungey2018strategic}
\bibinfo{author}{Dungey, M.H.}, \bibinfo{author}{Ghahremanlou, A.},
  \bibinfo{author}{Long, N.V.}, \bibinfo{year}{2018}.
\newblock \bibinfo{title}{Strategic bidding of electric power generating
  companies: Evidence from the {Australian National Energy Market}}.
\newblock \bibinfo{journal}{CESifo Working Paper Series No. 6819, Available at
  SSRN} \DOIprefix\doi{https://ssrn.com/abstract=3126673}.
\bibitem[{Eichler et~al.(2014)Eichler, Grothe, Manner and
  Tuerk}]{eichler2014models}
\bibinfo{author}{Eichler, M.}, \bibinfo{author}{Grothe, O.},
  \bibinfo{author}{Manner, H.}, \bibinfo{author}{Tuerk, D.},
  \bibinfo{year}{2014}.
\newblock \bibinfo{title}{Models for short-term forecasting of spike
  occurrences in {Australian} electricity markets: a comparative study}.
\newblock \bibinfo{journal}{Journal of Energy Markets} \bibinfo{volume}{7},
  \bibinfo{pages}{311--317}.
\bibitem[{Fanone et~al.(2013)Fanone, Gamba and Prokopczuk}]{fanone2013case}
\bibinfo{author}{Fanone, E.}, \bibinfo{author}{Gamba, A.},
  \bibinfo{author}{Prokopczuk, M.}, \bibinfo{year}{2013}.
\newblock \bibinfo{title}{The case of negative day-ahead electricity prices}.
\newblock \bibinfo{journal}{Energy Economics} \bibinfo{volume}{35},
  \bibinfo{pages}{22--34}.
\bibitem[{Frolova(2016)}]{frolova}
\bibinfo{author}{Frolova, N.}, \bibinfo{year}{2016}.
\newblock \bibinfo{title}{{Estimation of Extreme Value Dependence: Application
  to Australian Spot Electricity Prices}}.
\newblock \bibinfo{journal}{Master Thesis, Department of Mathematical and
  Statistical Sciences, University of Alberta}
  \DOIprefix\doi{https://doi.org/10.7939/R34J0B53F}.
\bibitem[{Garnaut(2011)}]{garnaut11}
\bibinfo{author}{Garnaut, R.}, \bibinfo{year}{2011}.
\newblock \bibinfo{title}{Garnaut climate change review update paper 8:
  Transforming the electricity sector}.
\newblock \URLprefix
  \url{http://www.garnautreview.org.au/update-2011/update-papers/up8-transforming-the-electricity-sector.html}.
  \bibinfo{note}{accessed 1 December 2019}.
\bibitem[{Haldrup and Nielsen(2006)}]{haldrup06}
\bibinfo{author}{Haldrup, N.}, \bibinfo{author}{Nielsen, M.},
  \bibinfo{year}{2006}.
\newblock \bibinfo{title}{A regime switching long memory model for electricity
  prices}.
\newblock \bibinfo{journal}{Journal of Econometrics} \bibinfo{volume}{135},
  \bibinfo{pages}{349--376}.
\bibitem[{Han et~al.(2016)Han, Linton, Oka and Whang}]{han2016cross}
\bibinfo{author}{Han, H.}, \bibinfo{author}{Linton, O.}, \bibinfo{author}{Oka,
  T.}, \bibinfo{author}{Whang, Y.J.}, \bibinfo{year}{2016}.
\newblock \bibinfo{title}{The cross-quantilogram: measuring quantile dependence
  and testing directional predictability between time series}.
\newblock \bibinfo{journal}{Journal of Econometrics} \bibinfo{volume}{193},
  \bibinfo{pages}{251--270}.
\bibitem[{Han et~al.(2020)Han, Kordzakhia and Tr{\"u}ck}]{han2020volatility}
\bibinfo{author}{Han, L.}, \bibinfo{author}{Kordzakhia, N.},
  \bibinfo{author}{Tr{\"u}ck, S.}, \bibinfo{year}{2020}.
\newblock \bibinfo{title}{Volatility spillovers in {A}ustralian electricity
  markets}.
\newblock \bibinfo{journal}{Energy Economics} \bibinfo{volume}{90},
  \bibinfo{pages}{104782}.
\bibitem[{Harris(2011)}]{harris2011electricity}
\bibinfo{author}{Harris, C.}, \bibinfo{year}{2011}.
\newblock \bibinfo{title}{Electricity markets: pricing, structures and
  economics}. volume \bibinfo{volume}{565}.
\newblock \bibinfo{publisher}{John Wiley \& Sons}.
\bibitem[{Hautsch and Herrera(2020)}]{hautsch2020multivariate}
\bibinfo{author}{Hautsch, N.}, \bibinfo{author}{Herrera, R.},
  \bibinfo{year}{2020}.
\newblock \bibinfo{title}{Multivariate dynamic intensity peaks-over-threshold
  models}.
\newblock \bibinfo{journal}{Journal of Applied Econometrics}
  \bibinfo{volume}{35}, \bibinfo{pages}{248--272}.
\bibitem[{Herrera and Clements(2018)}]{herrera2018point}
\bibinfo{author}{Herrera, R.}, \bibinfo{author}{Clements, A.},
  \bibinfo{year}{2018}.
\newblock \bibinfo{title}{Point process models for extreme returns: Harnessing
  implied volatility}.
\newblock \bibinfo{journal}{Journal of Banking \& Finance}
  \bibinfo{volume}{88}, \bibinfo{pages}{161--175}.
\bibitem[{Herrera and Gonz{\'a}lez(2014)}]{herrera2014modeling}
\bibinfo{author}{Herrera, R.}, \bibinfo{author}{Gonz{\'a}lez, N.},
  \bibinfo{year}{2014}.
\newblock \bibinfo{title}{The modeling and forecasting of extreme events in
  electricity spot markets}.
\newblock \bibinfo{journal}{International Journal of Forecasting}
  \bibinfo{volume}{30}, \bibinfo{pages}{477--490}.
\bibitem[{Higgs(2009)}]{HiggsH2009}
\bibinfo{author}{Higgs, H.}, \bibinfo{year}{2009}.
\newblock \bibinfo{title}{Modelling price and volatility inter-relationships in
  the {Australian} wholesale spot electricity markets}.
\newblock \bibinfo{journal}{Energy Economics} \bibinfo{volume}{31(5)},
  \bibinfo{pages}{748--756}.
\bibitem[{Hurn et~al.(2016)Hurn, Silvennoinen and
  Ter{\"a}svirta}]{hurn2016smooth}
\bibinfo{author}{Hurn, A.S.}, \bibinfo{author}{Silvennoinen, A.},
  \bibinfo{author}{Ter{\"a}svirta, T.}, \bibinfo{year}{2016}.
\newblock \bibinfo{title}{A smooth transition logit model of the effects of
  deregulation in the electricity market}.
\newblock \bibinfo{journal}{Journal of Applied Econometrics}
  \bibinfo{volume}{31}, \bibinfo{pages}{707--733}.
\bibitem[{Ignatieva and Tr\"{u}ck(2016)}]{IgnTr:2016}
\bibinfo{author}{Ignatieva, K.}, \bibinfo{author}{Tr\"{u}ck, S.},
  \bibinfo{year}{2016}.
\newblock \bibinfo{title}{Modeling spot price dependence in {Australian}
  electricity markets with applications to risk management}.
\newblock \bibinfo{journal}{Computers and Operations Research}
  \bibinfo{volume}{66}, \bibinfo{pages}{415--433}.
\bibitem[{Korniichuk(2012)}]{korniichuk2012forecasting}
\bibinfo{author}{Korniichuk, V.}, \bibinfo{year}{2012}.
\newblock \bibinfo{title}{Forecasting extreme electricity spot prices}.
\newblock \bibinfo{journal}{Cologne Graduate School Working Paper Series}
  \bibinfo{volume}{3}, \bibinfo{pages}{1--20}.
\bibitem[{Le~Pen and S\'{e}vi(2010)}]{lepen08}
\bibinfo{author}{Le~Pen, Y.}, \bibinfo{author}{S\'{e}vi, B.},
  \bibinfo{year}{2010}.
\newblock \bibinfo{title}{Volatility transmission and volatility impulse
  response functions in {E}uropean electricity forward markets}.
\newblock \bibinfo{journal}{Energy Economics} \bibinfo{volume}{32(4)},
  \bibinfo{pages}{758--770}.
\bibitem[{Li et~al.(2015)Li, Li and Tsai}]{li2015quantile}
\bibinfo{author}{Li, G.}, \bibinfo{author}{Li, Y.}, \bibinfo{author}{Tsai,
  C.L.}, \bibinfo{year}{2015}.
\newblock \bibinfo{title}{Quantile correlations and quantile autoregressive
  modeling}.
\newblock \bibinfo{journal}{Journal of the American Statistical Association}
  \bibinfo{volume}{110}, \bibinfo{pages}{246--261}.
\bibitem[{Linton and Whang(2007)}]{linton2007quantilogram}
\bibinfo{author}{Linton, O.}, \bibinfo{author}{Whang, Y.J.},
  \bibinfo{year}{2007}.
\newblock \bibinfo{title}{{The quantilogram: With an application to evaluating
  directional predictability}}.
\newblock \bibinfo{journal}{Journal of Econometrics} \bibinfo{volume}{141},
  \bibinfo{pages}{250--282}.
\bibitem[{Little(2013)}]{little2013oxford}
\bibinfo{author}{Little, T.D.}, \bibinfo{year}{2013}.
\newblock \bibinfo{title}{Time series analysis}, in: \bibinfo{booktitle}{The
  Oxford Handbook of Quantitative Methods in Psychology: Vol. 2}.
  \bibinfo{publisher}{Oxford University Press}. volume~\bibinfo{volume}{2}.
\bibitem[{Manner et~al.(2019)Manner, Fard, Pourkhanali and
  Tafakori}]{manner2019forecasting}
\bibinfo{author}{Manner, H.}, \bibinfo{author}{Fard, F.A.},
  \bibinfo{author}{Pourkhanali, A.}, \bibinfo{author}{Tafakori, L.},
  \bibinfo{year}{2019}.
\newblock \bibinfo{title}{Forecasting the joint distribution of {Australian}
  electricity prices using dynamic vine copulae}.
\newblock \bibinfo{journal}{Energy Economics} \bibinfo{volume}{78},
  \bibinfo{pages}{143--164}.
\bibitem[{Manner et~al.(2016)Manner, T{\"u}rk and Eichler}]{manner2016modeling}
\bibinfo{author}{Manner, H.}, \bibinfo{author}{T{\"u}rk, D.},
  \bibinfo{author}{Eichler, M.}, \bibinfo{year}{2016}.
\newblock \bibinfo{title}{Modeling and forecasting multivariate electricity
  price spikes}.
\newblock \bibinfo{journal}{Energy Economics} \bibinfo{volume}{60},
  \bibinfo{pages}{255--265}.
\bibitem[{Matsui and Mikosch(2016)}]{matsui2016extremogram}
\bibinfo{author}{Matsui, M.}, \bibinfo{author}{Mikosch, T.},
  \bibinfo{year}{2016}.
\newblock \bibinfo{title}{{The extremogram and the cross-extremogram for a
  bivariate GARCH (1, 1) process}}.
\newblock \bibinfo{journal}{Advances in Applied Probability}
  \bibinfo{volume}{48}, \bibinfo{pages}{217--233}.
\bibitem[{Mayer and Tr\"{u}ck(2018)}]{mayertrueck}
\bibinfo{author}{Mayer, K.}, \bibinfo{author}{Tr\"{u}ck, S.},
  \bibinfo{year}{2018}.
\newblock \bibinfo{title}{Electricity markets around the world}.
\newblock \bibinfo{journal}{Journal of Commodity Markets} \bibinfo{volume}{9},
  \bibinfo{pages}{77--100}.
\bibitem[{Mikosch and Zhao(2015)}]{mikosch2015integrated}
\bibinfo{author}{Mikosch, T.}, \bibinfo{author}{Zhao, Y.},
  \bibinfo{year}{2015}.
\newblock \bibinfo{title}{The integrated periodogram of a dependent extremal
  event sequence}.
\newblock \bibinfo{journal}{Stochastic Processes and their Applications}
  \bibinfo{volume}{125}, \bibinfo{pages}{3126--3169}.
\bibitem[{Muniain and Ziel(2020)}]{muniain2020probabilistic}
\bibinfo{author}{Muniain, P.}, \bibinfo{author}{Ziel, F.},
  \bibinfo{year}{2020}.
\newblock \bibinfo{title}{{Probabilistic forecasting in day-ahead electricity
  markets: Simulating peak and off-peak prices}}.
\newblock \bibinfo{journal}{International Journal of Forecasting}
  \bibinfo{volume}{In Press}.
\newblock
  \DOIprefix\doi{https://doi-org.simsrad.net.ocs.mq.edu.au/10.1016/j.ijforecast.2019.11.006}.
\bibitem[{Nepal et~al.(2016)Nepal, Foster et~al.}]{nepal2013testing}
\bibinfo{author}{Nepal, R.}, \bibinfo{author}{Foster, J.}, et~al.,
  \bibinfo{year}{2016}.
\newblock \bibinfo{title}{Testing for market integration in the {Australian
  National Electricity Market}}.
\newblock \bibinfo{journal}{The Energy Journal} \bibinfo{volume}{37},
  \bibinfo{pages}{215--238}.
\bibitem[{Park et~al.(2006)Park, Mjelde and Bessler}]{park06}
\bibinfo{author}{Park, H.}, \bibinfo{author}{Mjelde, J.},
  \bibinfo{author}{Bessler, D.}, \bibinfo{year}{2006}.
\newblock \bibinfo{title}{Price dynamics among {U.S.} markets}.
\newblock \bibinfo{journal}{Energy Economics} \bibinfo{volume}{28(1)},
  \bibinfo{pages}{81--101}.
\bibitem[{{Productivity Commission}(2013)}]{agpc13}
\bibinfo{author}{{Productivity Commission}}, \bibinfo{year}{2013}.
\newblock \bibinfo{title}{Electricity network regulatory frameworks,
  {Productivity Commission} inquiry report}.
\newblock \URLprefix
  \url{http://www.pc.gov.au/inquiries/completed/electricity/report}.
  \bibinfo{note}{accessed 30 April 2016}.
\bibitem[{Rai and Nelson(2019)}]{rai2019australia}
\bibinfo{author}{Rai, A.}, \bibinfo{author}{Nelson, T.}, \bibinfo{year}{2019}.
\newblock \bibinfo{title}{{Australia's National Electricity Market} after
  twenty years}.
\newblock \bibinfo{journal}{Australian Economic Review} \bibinfo{volume}{53},
  \bibinfo{pages}{165--182}.
\bibitem[{Rinaldi et~al.(2018)Rinaldi, Djuraidah, Wigena, Mangku and
  Gunawan}]{rinaldi2018identification}
\bibinfo{author}{Rinaldi, A.}, \bibinfo{author}{Djuraidah, A.},
  \bibinfo{author}{Wigena, A.H.}, \bibinfo{author}{Mangku, I.W.},
  \bibinfo{author}{Gunawan, D.}, \bibinfo{year}{2018}.
\newblock \bibinfo{title}{Identification of extreme rainfall pattern using
  extremogram in {West Java}}, in: \bibinfo{booktitle}{IOP Conference Series:
  Earth and Environmental Science}, \bibinfo{organization}{IOP Publishing}. p.
  \bibinfo{pages}{012064}.
\bibitem[{Schmitt et~al.(2015)Schmitt, Sch{\"a}fer, Dette and
  Guhr}]{schmitt2015quantile}
\bibinfo{author}{Schmitt, T.A.}, \bibinfo{author}{Sch{\"a}fer, R.},
  \bibinfo{author}{Dette, H.}, \bibinfo{author}{Guhr, T.},
  \bibinfo{year}{2015}.
\newblock \bibinfo{title}{{Quantile correlations: Uncovering temporal
  dependencies in financial time series}}.
\newblock \bibinfo{journal}{International Journal of Theoretical and Applied
  Finance} \bibinfo{volume}{18}, \bibinfo{pages}{1550044}.
\bibitem[{Smith(2015)}]{smith2015copula}
\bibinfo{author}{Smith, M.S.}, \bibinfo{year}{2015}.
\newblock \bibinfo{title}{Copula modelling of dependence in multivariate time
  series}.
\newblock \bibinfo{journal}{International Journal of Forecasting}
  \bibinfo{volume}{31}, \bibinfo{pages}{815--833}.
\bibitem[{Smith et~al.(2012)Smith, Gan and Kohn}]{smith2010}
\bibinfo{author}{Smith, M.S.}, \bibinfo{author}{Gan, Q.},
  \bibinfo{author}{Kohn, R.J.}, \bibinfo{year}{2012}.
\newblock \bibinfo{title}{{Modelling dependence using skew t copulas: Bayesian
  inference and applications}}.
\newblock \bibinfo{journal}{Journal of Applied Econometrics}
  \bibinfo{volume}{27}, \bibinfo{pages}{500--522}.
\bibitem[{Smith and Shively(2018)}]{smith2018econometric}
\bibinfo{author}{Smith, M.S.}, \bibinfo{author}{Shively, T.S.},
  \bibinfo{year}{2018}.
\newblock \bibinfo{title}{Econometric modeling of regional electricity spot
  prices in the {Australian} market}.
\newblock \bibinfo{journal}{Energy Economics} \bibinfo{volume}{74},
  \bibinfo{pages}{886--903}.
\bibitem[{Todorova(2017)}]{todorova2017intraday}
\bibinfo{author}{Todorova, N.}, \bibinfo{year}{2017}.
\newblock \bibinfo{title}{{The intraday directional predictability of large
  Australian stocks: A cross-quantilogram analysis}}.
\newblock \bibinfo{journal}{Economic Modelling} \bibinfo{volume}{64},
  \bibinfo{pages}{221--230}.
\bibitem[{Weron(2006)}]{Weron2006}
\bibinfo{author}{Weron, R.}, \bibinfo{year}{2006}.
\newblock \bibinfo{title}{Modeling and Forecasting Loads and Prices in
  Deregulated Electricity Markets}.
\newblock \bibinfo{publisher}{Wiley, Chichester}.
\bibitem[{White et~al.(2015)White, Kim and Manganelli}]{white2015var}
\bibinfo{author}{White, H.}, \bibinfo{author}{Kim, T.H.},
  \bibinfo{author}{Manganelli, S.}, \bibinfo{year}{2015}.
\newblock \bibinfo{title}{{VAR for VaR: Measuring tail dependence using
  multivariate regression quantiles}}.
\newblock \bibinfo{journal}{Journal of Econometrics} \bibinfo{volume}{187},
  \bibinfo{pages}{169--188}.
\bibitem[{Wood and Blowers(2018)}]{wood2018mostly}
\bibinfo{author}{Wood, A.}, \bibinfo{author}{Blowers, D.},
  \bibinfo{year}{2018}.
\newblock \bibinfo{title}{Mostly Working: {Australia's} Wholesale Electricity
  Market}.
\newblock \bibinfo{publisher}{Grattan Institute}.
\bibitem[{Worthington et~al.(2005)Worthington, Kay-Spratley and
  Higgs}]{WKH2005}
\bibinfo{author}{Worthington, A.}, \bibinfo{author}{Kay-Spratley, A.},
  \bibinfo{author}{Higgs, H.}, \bibinfo{year}{2005}.
\newblock \bibinfo{title}{Transmission of prices and price volatility in
  {A}ustralian electricity spot markets: A multivariate {GARCH} analysis}.
\newblock \bibinfo{journal}{Energy Economics} \bibinfo{volume}{27(2)},
  \bibinfo{pages}{337--350}.
\bibitem[{Yan and Trück(2020)}]{yan2020}
\bibinfo{author}{Yan, G.}, \bibinfo{author}{Trück, S.}, \bibinfo{year}{2020}.
\newblock \bibinfo{title}{A dynamic network analysis of spot electricity prices
  in the {A}ustralian national electricity market}.
\newblock \bibinfo{journal}{Energy Economics} \bibinfo{volume}{92},
  \bibinfo{pages}{104972}.
\bibitem[{Zachmann(2008)}]{zachmann08}
\bibinfo{author}{Zachmann, G.}, \bibinfo{year}{2008}.
\newblock \bibinfo{title}{Electricity wholesale market prices in {E}urope:
  {C}onvergence?}
\newblock \bibinfo{journal}{Energy Economics} \bibinfo{volume}{30(4)},
  \bibinfo{pages}{1659--1671}.

\end{thebibliography}

\newpage
\appendix
	\section{Spot prices in the NEM markets}

	\setcounter{figure}{0}
	
	\begin{figure}[!htbp]
		\centering
		\includegraphics[width=1\textwidth]{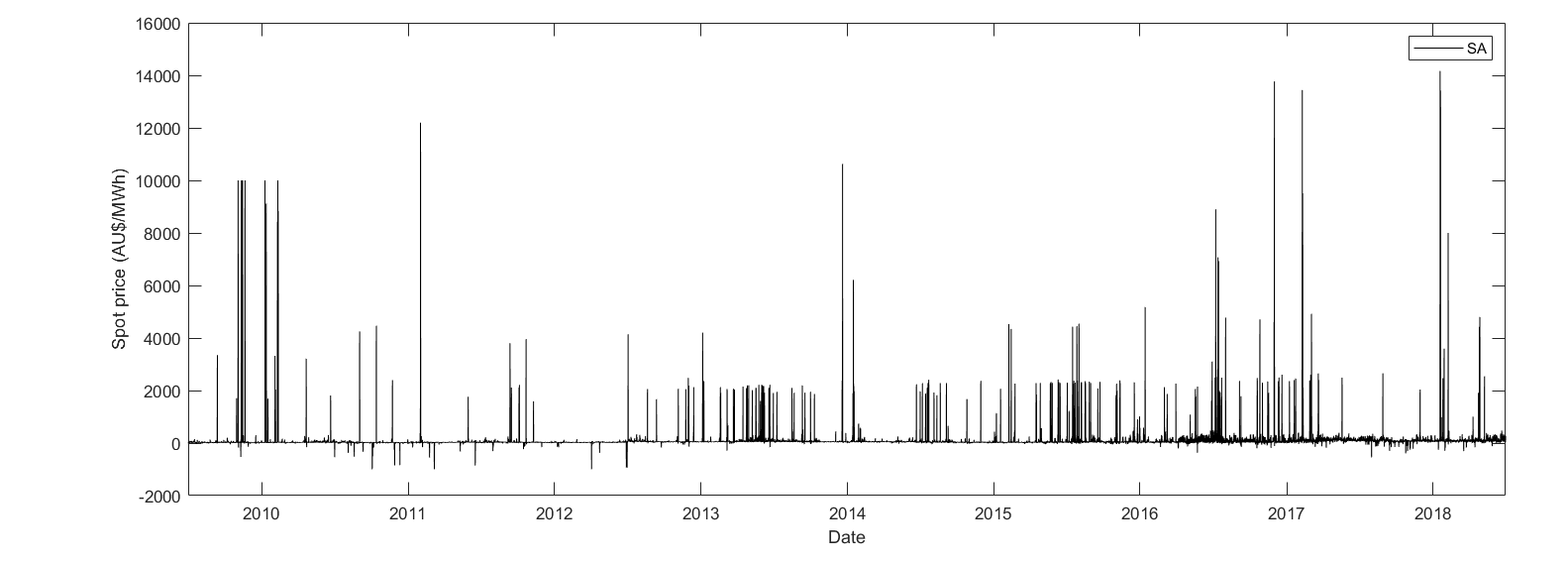}
		\includegraphics[width=1\textwidth]{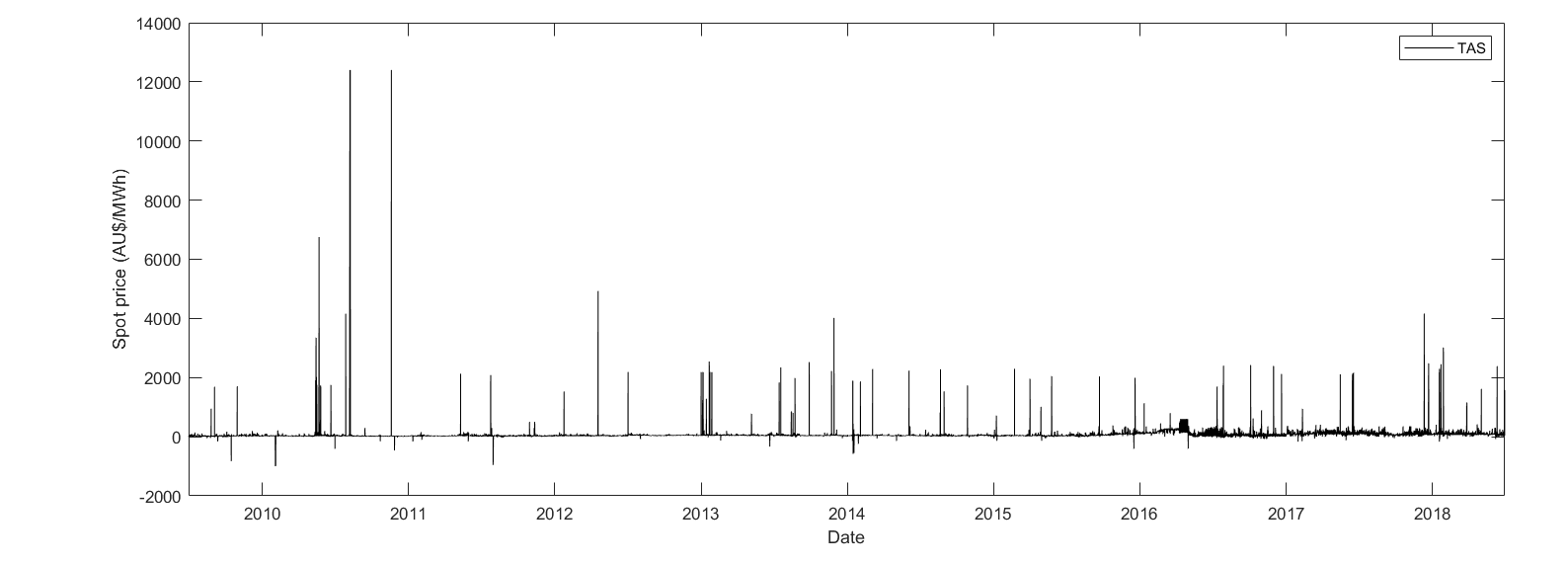}
		\includegraphics[width=1\textwidth]{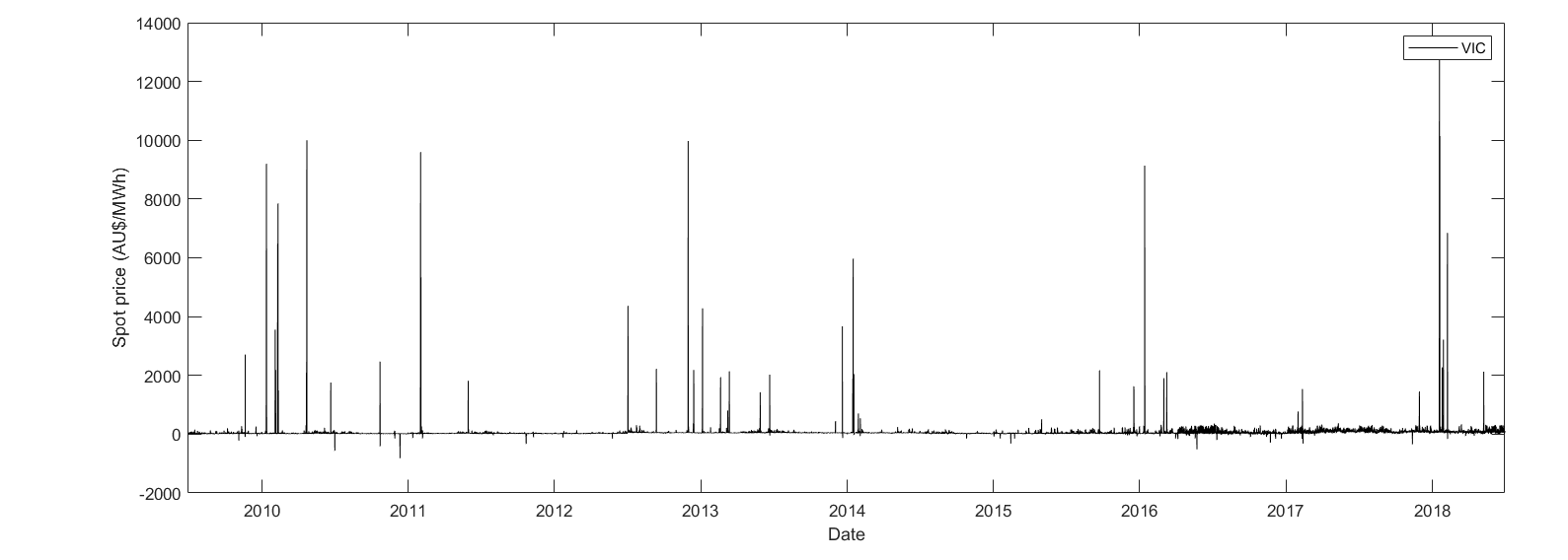}
		\caption{Half-hourly spot prices (A\$/MWh) for the SA, TAS and VIC electricity markets from July 1, 2009 to June 30, 2018.}
		\label{Afig:price plot}
	\end{figure}

\newpage

\section{Cross-Extremograms for 5-minute prices}

\setcounter{figure}{0}

\begin{figure}[!htbp]
	\centering
	\includegraphics[width=\linewidth,height=15cm]{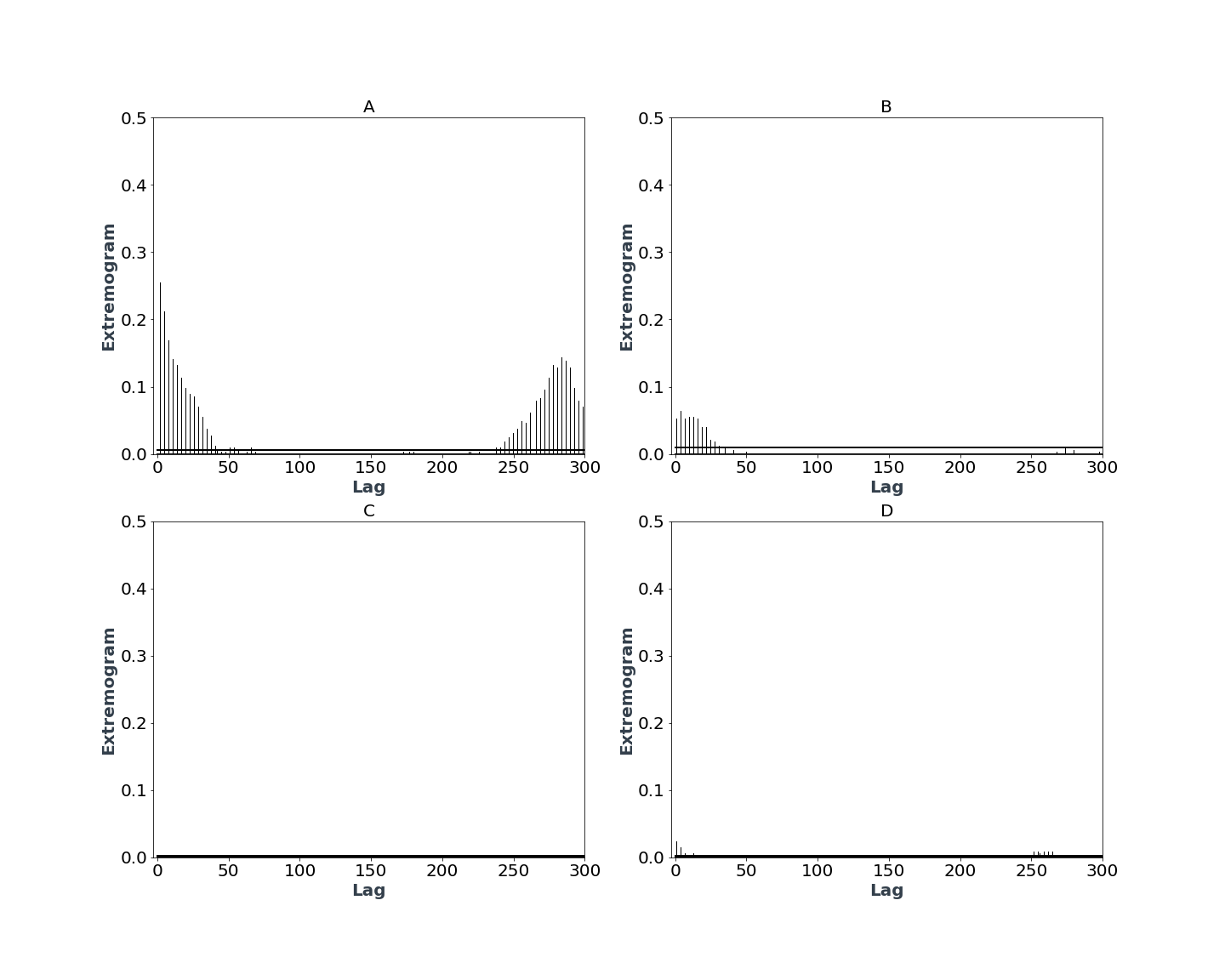}
	\caption[Cross-extremograms (5-minute) conditioning on price spikes in NSW]{Sample cross-extremograms of 5-minute spot electricity prices conditioning on price spikes in NSW. The plot provides cross-extremograms for (A) NSW $\rightarrow$ QLD, (B) NSW $\rightarrow$ SA, (C) NSW $\rightarrow$ TAS and (D) NSW $\rightarrow$ VIC based on a sample period from 1 July 2009 to 30 June 2018.}
	\label{fig:cross_ext_NSW_5min}
\end{figure}

\begin{figure}[!htbp]
	\centering
	\includegraphics[width=\linewidth,height=15cm]{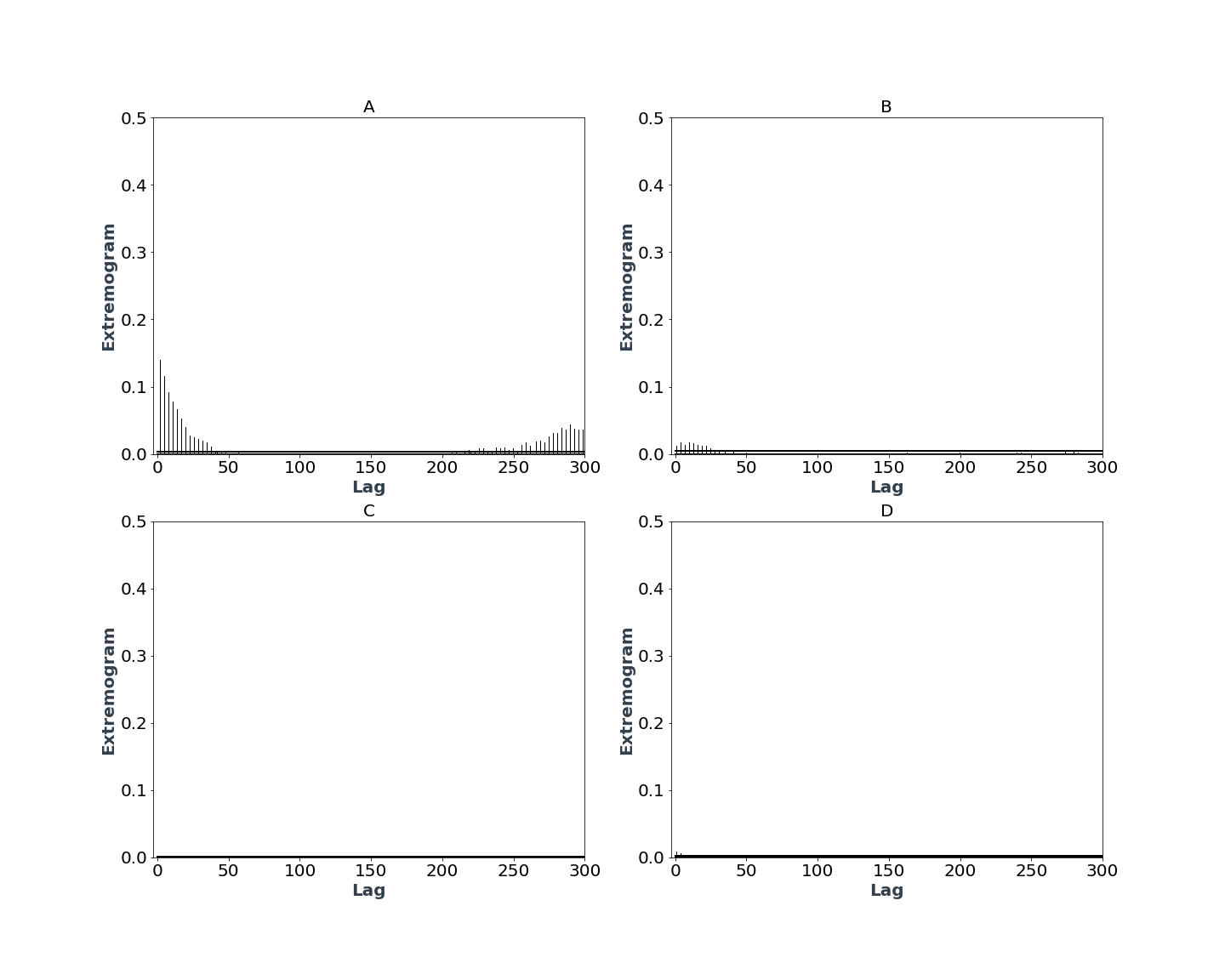}
	\caption[Cross-extremograms (5-minute) conditioning on price spikes in QLD]{Sample cross-extremograms of 5-minute spot electricity prices conditioning on price spikes in QLD. The plot provides cross-extremograms for (A) QLD $\rightarrow$ NSW, (B) QLD $\rightarrow$ SA, (C) QLD $\rightarrow$ TAS and (D) QLD $\rightarrow$ VIC based on a sample period from 1 July 2009 to 30 June 2018.}
	\label{fig:cross_ext_QLD_5min}
\end{figure}

\begin{figure}[!htbp]
	\centering
	\includegraphics[width=\linewidth,height=15cm]{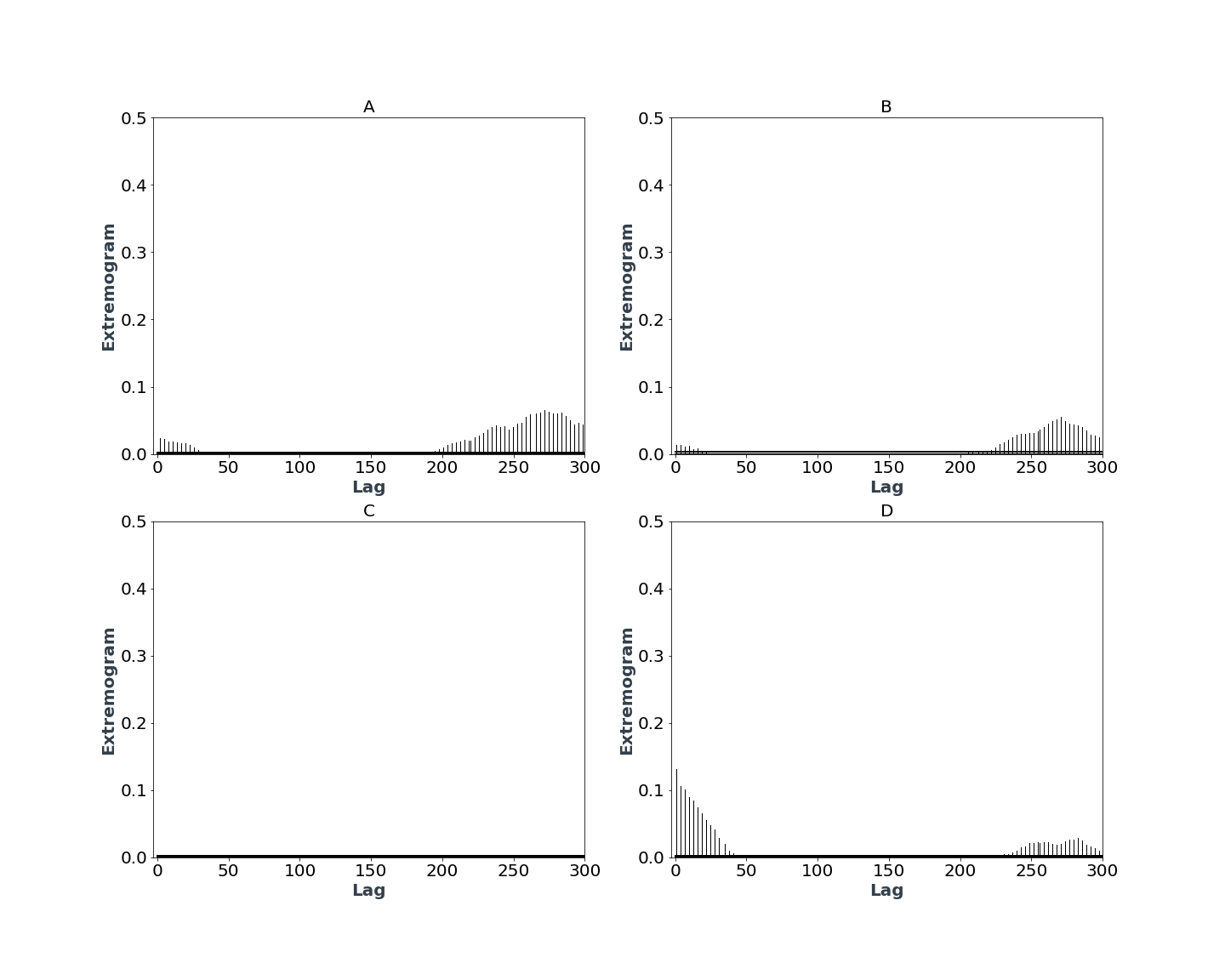}
	\caption[Cross-extremograms (5-minute) conditioning on price spikes in SA]{Sample cross-extremograms of 5-minute spot electricity prices conditioning on price spikes in SA. The plot provides cross-extremograms for (A) SA $\rightarrow$ NSW, (B) SA $\rightarrow$ QLD, (C) SA $\rightarrow$ TAS and (D) SA $\rightarrow$ VIC based on a sample period from 1 July 2009 to 30 June 2018.}
	\label{fig:cross_ext_SA_5min}
\end{figure}

\begin{figure}[!htbp]
	\centering
	\includegraphics[width=\linewidth,height=15cm]{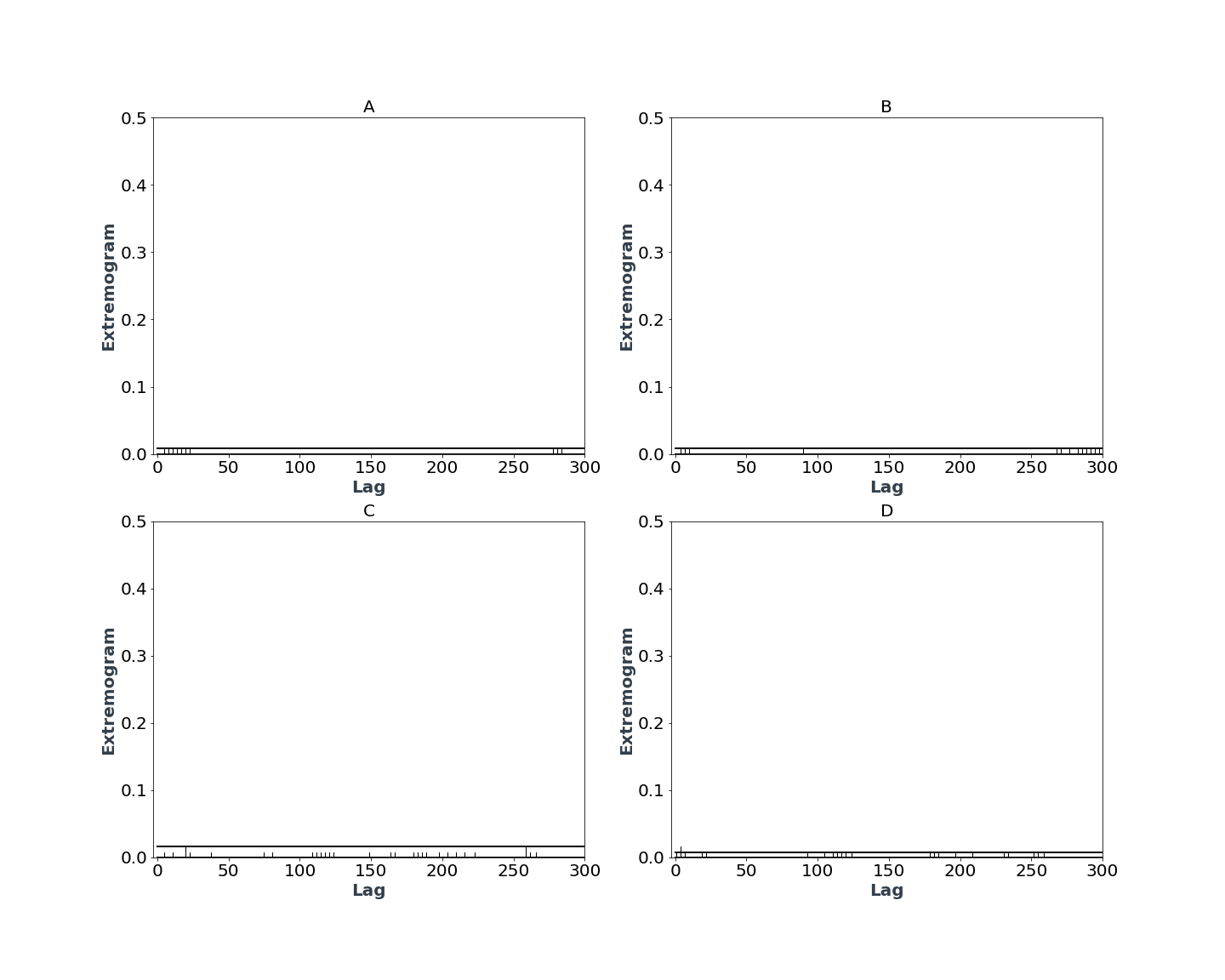}
	\caption[Cross-extremograms (5-minute) conditioning on price spikes in TAS]{Sample cross-extremograms of 5-minute spot electricity prices conditioning on price spikes in TAS. The plot provides cross-extremograms for (A) TAS $\rightarrow$ NSW, (B) TAS $\rightarrow$ QLD, (C) TAS $\rightarrow$ SA and (D) TAS $\rightarrow$ VIC based on a sample period from 1 July 2009 to 30 June 2018.}
	\label{fig:cross_ext_TAS_5min}
\end{figure}

\begin{figure}[!htbp]
	\centering
	\includegraphics[width=\linewidth,height=15cm]{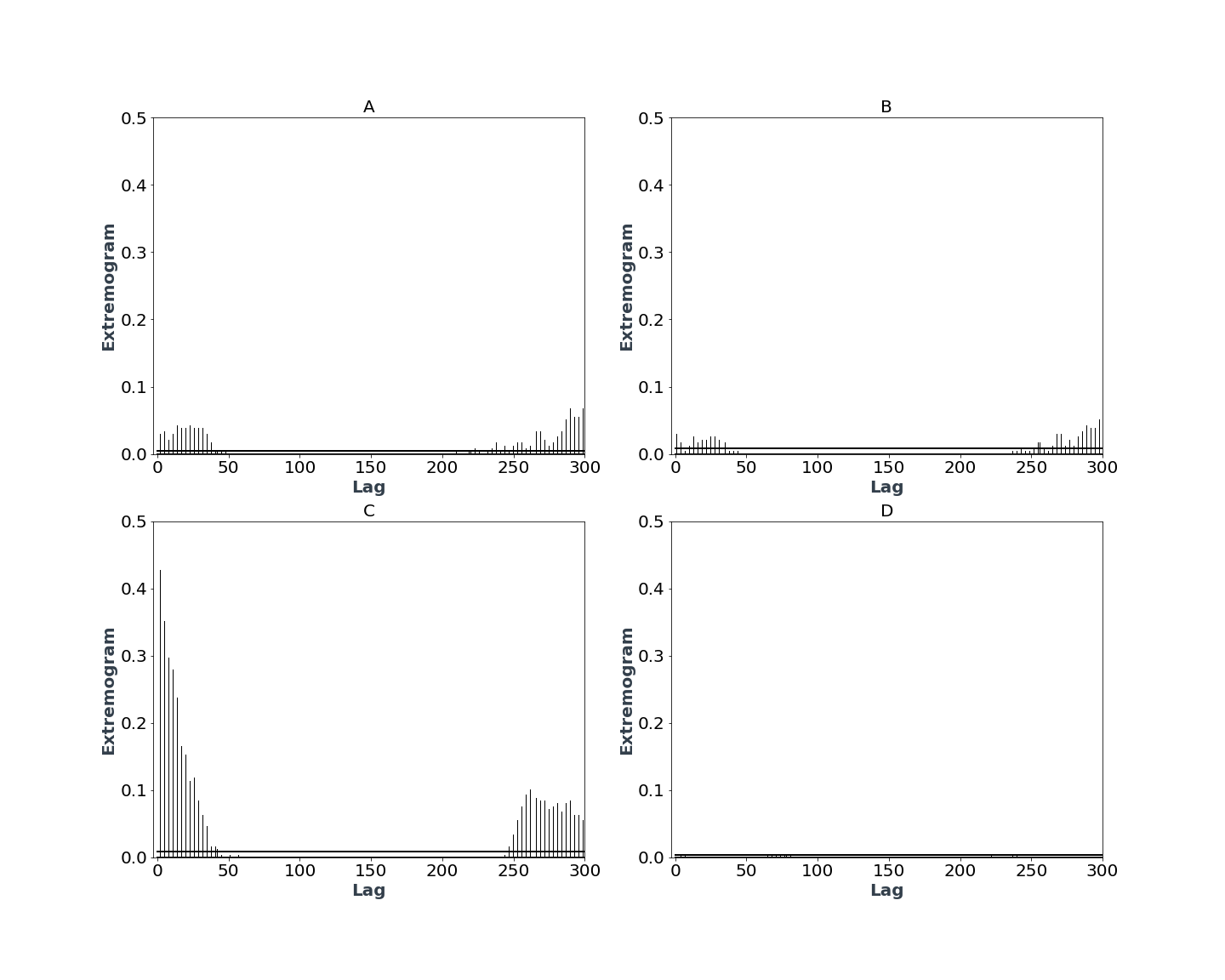}
	\caption[Cross-extremograms (5-minute) conditioning on price spikes in VIC]{Sample cross-extremograms of 5-minute spot electricity prices conditioning on price spikes in VIC. The plot provides cross-extremograms for (A) VIC $\rightarrow$ NSW, (B) VIC $\rightarrow$ QLD, (C) VIC $\rightarrow$ SA and (D) VIC $\rightarrow$ TAS based on a sample period from 1 July 2009 to 30 June 2018.}
	\label{fig:cross_ext_VIC_5min}
\end{figure}
	
\end{document}